\documentclass{svmult}

\title*{
Optimal consumption and investment with bounded downside risk
for power utility functions}

\titlerunning{Consumption and investment with bounded downside risk}
\author{Claudia Kl\"uppelberg and Serguei Pergamenchtchikov}
\institute{
Claudia Kl\"uppelberg \at
Center for Mathematical Sciences,
Technische Universit\"at M\"unchen, D-85747 Garching, Germany,
\email{cklu@ma.tum.de}
\and Serguei Pergamenchtchikov \at
Laboratoire de Math\'ematiques Rapha\"el Salem,
 Universit\'e de Rouen,   BP.12,
76801 Saint Etienne du Rouvray, France,
\email{Serge.Pergamenchtchikov@univ-rouen.fr}  }

\usepackage{mathptmx}       
\usepackage{helvet}         
\usepackage{courier}        
\usepackage{type1cm}        
\usepackage{type1cm}
\usepackage{amssymb}
\usepackage{amsfonts}
\usepackage{amsmath}

\usepackage{makeidx}         
\usepackage{graphicx}        
\usepackage{multicol}        
\usepackage[bottom]{footmisc}

\newtheorem{conclusion}[theorem]{Conclusion}

\newcommand{\beao}{\begin{eqnarray*}}
\newcommand{\eeao}{\end{eqnarray*}\noindent}

\newcommand{\beam}{\begin{eqnarray}}
\newcommand{\eeam}{\end{eqnarray}\noindent}

\def\bbr{{\mathbb R}}
\def\bbn{{\mathbb N}}

\newcommand{\stp}{\stackrel{P}{\rightarrow}}

\newcommand{\stas}{\stackrel{\rm a.s.}{\rightarrow}}

\newcommand{\nto}{n\to\infty}

\newcommand{\al}{{\alpha}}
\newcommand{\la}{{\lambda}}

\newcommand{\ga}{{\gamma}}
\newcommand{\si}{{\sigma}}

\newcommand{\wh}{\widehat}
\newcommand{\wt}{\widetilde}

\newcommand{\eproof}{\quad\hfill\mbox{$\Box$}\\[3mm]}

\newcommand\cE{{\cal E}}

\newcommand\cF{{\cal F}}

\newcommand\cB{{\cal B}}
\newcommand\cK{{\cal K}}

\newcommand\cU{{\cal U}}
\newcommand\cV{{\cal V}}

\def\text#1{\hbox{#1}}
\def\proof{{\noindent \bf Proof. }}

\def\eproof{\quad\hfill\mbox{$\Box$}}
\newcommand{\halmos}{\quad\hfill\mbox{$\Box$}}

\def\E{{\bf E}}
\def\P{{\bf P}}

\def\H{{\bf H}}

\def\Chi{{\bf 1}}
\def\d{{\rm d}}
\def\build #1_#2{\mathrel{\mathop{\kern 0pt #1}\limits_\zs{#2}}}
\newcommand{\zs}[1]{{\mathchoice{#1}{#1}{\lower.25ex\hbox{$\scriptstyle#1$}}
{\lower0.25ex\hbox{$\scriptscriptstyle#1$}}}}

\begin{document}

\maketitle

\abstract{We investigate optimal consumption and investment problems for a Black-Scholes market
under uniform restrictions on Value-at-Risk and Expected Shortfall.
We formulate various utility maximization problems, which can be solved
explicitly. We compare the optimal solutions in form of
optimal value, optimal control and optimal wealth to analogous
problems under additional uniform risk bounds.
Our proofs are partly based on solutions to Hamilton-Jacobi-Bellman
equations, and we prove a corresponding verification theorem.
\footnotetext{This work was supported by the European Science Foundation
through the AMaMeF programme.}
}


\vspace*{2mm}


\section{Introduction}\label{sec:1}

We consider an investment problem aiming at optimal consumption
during a fixed  investment
interval $[0,T]$ in addition to
an optimal terminal wealth at maturity $T$.
Such problems are of prime interest for the institutional
investor, selling asset funds to their customers, who are
entitled to certain payment during the duration of an investment contract
and expect a high return at maturity.
The classical approach to this problem goes back to Merton~\cite{Me} and involves utility functions,
more precisely, the expected utility serves as the functional
which has to be optimized.

We adapt this classical utility maximization approach to today's industry practice:
investment firms customarily impose limits on the risk of trading portfolios.
These limits are specified in terms of downside risk measures as the popular Value-at-Risk (VaR) or  Expected Shortfall (ES). We briefly comment on these two risk measures.

As Jorion~\cite{Jo}, p. 379  points out, VaR creates a common denominator for the comparison of different risk activities. Traditionally, position limits of traders are set in terms of notional exposure, which may not be directly comparable across treasuries with different maturities.
In contrast, VaR provides a common denominator to compare various asset classes and business units.
The popularity of VaR as a risk measure has been endorsed by regulators, in particular, the Basel Committee on Banking Supervision, which resulted in mandatory regulations worldwide.
One of the well-known drawbacks of VaR is due to its definition as a quantile.
This means that only the probability to exceed a VaR bound is considered,
the values of the losses are not taken into account.
Artzner et al.~\cite{ArDeEbHe} proposes as an alternative risk measure the Expected Shortfall,
defined as the conditional expectation of losses above VaR.

Our approach combines the classical utility maximization with risk limits in terms of VaR and ES.
This leads to control problems under restrictions on uniform versions of VaR
or ES, where the risk bound is supposed to be
in vigour throughout the duration of the investment.
To our knowledge such problems have only been considered in dynamic settings which reduce intrinsically to static problems.
Emmer, Kl\"uppelberg and
Korn~\cite{EmKlKo} consider a dynamic market, but maximize only the expected wealth at maturity
under a downside risk bound at maturity.
Basak and Shapiro \cite{BaSha} solve the utility optimization problem for complete markets with bounded VaR at maturity.
Gabih, Gretsch and Wunderlich
\cite{GaGrWu} solve the utility optimization problem
for constant coefficients markets with bounded ES at maturity.

In the present paper we aim now at a truly dynamic portfolio choice of a
trader subject to a risk limit specified in terms of VaR or ES.
We shall start with Merton's consumption and investment problem for a pricing model driven by
Brownian motion with c\`adl\`ag drift and volatility coefficients.
Such dynamic optimization problems for standard financial markets have been solved in Karatzas and Shreve~\cite{KaSh2} by martingale methods.
In order to obtain the optimal strategy in ``feedback form'' basic assumption in \cite{KaSh2} on the coefficients is H\"older continuity of a certain order (see e.g. Assumption 8.1, p. 119).
In the present paper we use classical optimization methods from stochastic control. This makes it possible to formulate optimal solutions to
Merton's consumption and investment problem in ``explicit feedback form'' for different power consumption and wealth utility functions. We also weaken the H\"older continuity assumption to  c\`adl\`ag coefficients satisfying weak integrability conditions.

In a second step we introduce uniform risk limits in terms of VaR and ES into this optimal consumption and investment problem.
Our risk measures are specified to represent the required Capital-at-Risk of the institutional investor.
The amount of required capital increases with the corresponding loss quantile
representing the security of the investment.
This quantile is for any specific trader an exogeneous variable, which he/she cannot influence.
Additionally, each trader can set a specific portfolio's risk limit, which may affect the already exogeneously given risk limit of the portfolio.
A trader, who has been given a fixed Capital-at-Risk, can now
use risk limits for different portfolios categorizing
the riskiness of his/her portfolios in this way.

It has been observed by Basak and Shapiro \cite{BaSha}
that VaR limits only applied at maturity can actually increase the risk.
In contrast to this observation, when working with a power utility function
and a uniform risk limit throughout the investment horizon, this effect disappears;  indeed the optimal strategy for the constrained problem
of Theorem~\ref{Th.3.12} given in \eqref{3.27} is riskless for sufficiently small risk bound:
For a HARA utility function, in order to keep within a sufficiently small risk bound, it is not allowed to invest anything into risky assets at all, but consume everything. This is in contrast to the optimal strategy, when we optimise the linear utility, which recommends to invest everything into risky assets and consume nothing; see \eqref{3.12} of Theorem~{Th.3.1}

Within the class of admissible control processes we identify subclasses of controls,
which allow for an explicit expression of the optimal strategy.
We derive results based on certain utility maximization strategies, choosing a power
utility function for both, the consumption process and the terminal wealth.
The literature to utility maximization is vast;
we only mention the books by
Karatzas and Shreve~\cite{KaSh1,KaSh2}, Korn~\cite{Ko} and Merton~\cite{Me}.
Usually, utility maximization is based on concave utility functions.
The assumption of concavity models the idea that the infinitesimal
utility decreases with increasing wealth.
Within the class of power utility functions this corresponds to parameters $\gamma<1$.
The case $\gamma=1$ corresponds to  linear utility functions,
meaning that expected utility
reduces to expected wealth.

Our paper is organised as follows.
In Section~\ref{sec:2} we formulate the problem.
In Section~\ref{subsec:2.1} the Black-Scholes model for the price processes
and the  parameter restrictions are presented. We also define the necessary quantities
like consumption and portfolio processes, also recall the notion of a self-financing
portfolio and a trading strategy. Section~\ref{subsec:2.2} is devoted
to the control processes; here
also the different classes of controls to be considered later are introduced.
The cost functions are defined in Section~\ref{subsec:2.3} and the risk measures in
Section~\ref{subsec:2.4}.
In Section~\ref{sec:3} all optimization problems and their solutions are given.
Here also the consequences for the trader are discussed.
All proofs are summarized in Section~\ref{sec:4} with a verification theorem postponed to
the Appendix.

\renewcommand{\theequation}{\arabic{section}.\arabic{equation}}
\section{Formulating the Problem}\label{sec:2}
\subsection{The Model}\label{subsec:2.1}
\setcounter{equation}{0}

We consider a Black-Scholes type financial market consisting of one
{\em riskless bond} and several {\em risky stocks}.
Their respective prices $(S_\zs{0}(t))_\zs{0\le t\le T}$ and
$(S_\zs{i}(t))_\zs{0\le t\le T}$ for $i=1,\ldots,d$
evolve according to the equations:
\begin{equation}\label{2.1}
\left\{\begin{array}{ll}
\d S_\zs{0}(t)\,=\,r_\zs{t}\,S_\zs{0}(t)\,\d t\,, & S_\zs{0}(0)\,=\,1\,,\\[5mm]
\d S_\zs{i}(t)\,=\,S_\zs{i}(t)\,\mu_\zs{i}(t)\,\d t\,+\,S_\zs{i}(t)\,
\sum^d_\zs{j=1}\,\sigma_\zs{ij}(t)\,\d W_\zs{j}(t)\,, &
S_\zs{i}(0)\,=\,s_i\,>0\,.
\end{array}\right.
\end{equation}
Here $W_\zs{t}=(W_\zs{1}(t),\ldots,W_\zs{d}(t))'$
is a standard $d$-dimensional Brownian motion;
$r_\zs{t}\in\bbr$ is the {\em riskless interest rate},
$\mu_\zs{t}=(\mu_\zs{1}(t),\ldots,\mu_\zs{d}(t))'\in\bbr^d$ is the vector of
{\em stock-appreciation rates} and
$\sigma_\zs{t}=(\sigma_\zs{ij}(t))_\zs{1\le i,j\le d}$ is
the matrix of {\em stock-volatilities}.
We assume that the coefficients $r_\zs{t}$, $\mu_\zs{t}$ and $\sigma_\zs{t}$
are deterministic functions, which are right continuous
with  left limits (c\`adl\`ag).
We also assume that the matrix $\sigma_\zs{t}$ is non-singular for
Lebesgue-almost all $t\ge0$.

We denote  by $\cF_\zs{t}=\sigma\{W_\zs{s}\,,s\le t \}$, $t\ge 0$, the filtration generated by the Brownian motion (augmented by the null sets).
Furthermore, $|\cdot|$ denotes the Euclidean norm for vectors and the
corresponding matrix norm for matrices.
For $(y_t)_\zs{0\le t\le T}$ square integrable over the fixed interval $[0,T]$ we define
$\|y\|_T=(\int_0^T |y_t|^2\,\d t)^{1/2}$.

For $t\ge 0$ let $\phi_t\in\bbr$ denote the amount of investment into bond and
$$
\varphi_\zs{t}=(\varphi_\zs{1}(t),\ldots,\varphi_\zs{d}(t))'\in\bbr^d
$$
the amount of investment into risky assets.
We recall  that a
{\em trading strategy} is an $\bbr^{d+1}$-valued $(\cF_\zs{t})_\zs{0\le t\le T}$-progressively measurable process
$(\phi_\zs{t},\varphi_\zs{t})_\zs{0\le t\le T}$ and that
$$
X_\zs{t}\,=\,\phi_\zs{t}\,S_\zs{0}(t)\,+
\sum^d_\zs{j=1}\,\varphi_\zs{j}(t)\,S_\zs{j}(t)\,,\quad t\ge 0\,,
$$
is called the {\em wealth process}.
Moreover, an $(\cF_\zs{t})_\zs{0\le t\le T}$-progressively measurable nonnegative
process $(c_\zs{t})_\zs{0\le t\le T}$ satisfying for the investment horizon $T>0$
$$
\int^T_0\,c_\zs{t}\,\d t\,<\,\infty \quad \mbox{a.s.}
$$
is called {\em consumption process}.

The trading strategy $((\phi_\zs{t},\varphi_\zs{t}))_\zs{0\le t\le T}$ and the consumption process
$(c_\zs{t})_\zs{0\le t\le T}$ are called {\em self-financing}, if the wealth process
 satisfies the following equation
\begin{equation}\label{2.2}
X_\zs{t}\,=\,x\,+\,
\int^t_0\,\phi_\zs{u}\,\d S_\zs{0}(u)\,+
\,\sum^d_\zs{j=1}\,\int^t_0\,\varphi_\zs{j}(u)\,\d S_\zs{j}(u)\,-\,
\int^t_0\,c_\zs{u}\,\d u\,, \quad t\ge 0\,,
\end{equation}
where $x>0$ is the initial endowment.

In this paper we work with relative quantities, i.e. with the fractions of the wealth process,
which are invested into bond and stocks; i.e., we define for $ j=1,\ldots,d$
$$
\pi_\zs{j}(t)\,:=\,\frac{\varphi_\zs{j}(t)\,S_\zs{j}(t)}
{\phi_\zs{t}\,S_\zs{0}(t)\,+\,
\sum^d_\zs{j=1}\,\varphi_\zs{i}(t)\,S_\zs{i}(t)}\,,\quad t\ge 0\,.
$$
Then
$\pi_\zs{t}=(\pi_\zs{1}(t),\ldots, \pi_\zs{d}(t))'$, $t\ge 0$, is called
the {\em portfolio process} and we assume throughout that
it is $(\cF_\zs{t})_\zs{0\le t\le T}$-progressively measurable.
We assume that for the fixed investment horizon $T>0$
$$
\|\pi\|_T^2:= \int_0^T |\pi_t|^2 \d t <\infty\quad {\rm a.s.}\,.
$$
We also define with $\Chi=(1,\ldots,1)'\in\bbr^d$ the quantities
\begin{equation}\label{2.3}
y_\zs{t}=\sigma'_\zs{t}\pi_\zs{t}\quad\mbox{and}\quad
\theta_\zs{t}=\sigma^{-1}_\zs{t}(\mu_\zs{t}-r_\zs{t}\,\Chi)\,,\quad t\ge 0\,,
\end{equation}
where it suffices that these quantities are defined for Lebesgue-almost all $t\ge 0$.
Taking these definitions into account we rewrite equation \eqref{2.2} for $X_\zs{t}$ as
\begin{equation}\label{2.4}
\d X_\zs{t}\,=\,X_\zs{t}\,(r_\zs{t}\,+\,y'_\zs{t}\,\theta_\zs{t})
\,\d t\,-\,c_\zs{t}\,\d t\,+\,X_\zs{t}\,y'_\zs{t}\,\d W_\zs{t}\,,\quad t>0\,,\quad
 X_\zs{0}\,=\,x>0\,.
\end{equation}
This implies in particular that any optimal investment strategy is equal to\\
$\pi^*_t=\sigma_t'^{-1}y^*_t$, where $y^*_t$ is the optimal control process
for equation \eqref{2.4}.
We also require for the investment horizon $T>0$
\begin{equation}\label{2.5}
\|\theta\|^2_\zs{T}=\int^T_\zs{0}\,|\theta_\zs{t}|^2\d t\,<\,\infty\,.
\end{equation}
Besides the already defined Euclidean norm we shall also use for arbitrary $q\ge 1$ the notation $\|f\|_\zs{q,T}$ for the $q$-norm of
$(f_\zs{t})$, i.e.
\begin{equation}\label{2.6}
\|f\|_\zs{q,T}=\left(\int^T_\zs{0}\,|f_\zs{t}|^q\d t\right)^{1/q}\,.
\end{equation}

\subsection{The Control Processes}
\label{subsec:2.2}

Now we introduce the set of control processes
$(y_\zs{t},c_\zs{t})_\zs{0\le t\le T}$.  First we choose the consumption process
$(c_\zs{t})_\zs{0\le t\le T}$ as a proportion of the wealth process; i.e.
$$
c_\zs{t}\,=\,v_\zs{t}\,X_\zs{t}\,,
$$
where $(v_t)_\zs{0\le t\le T}$
is a deterministic non-negative function satisfying
$$
\int^T_\zs{0}\,v_\zs{t}\,\d t\,<\,\infty\,.
$$
For this consumption we define the {\em control process}
$\varsigma=(\varsigma_\zs{t})_\zs{0\le t\le T}$
as
$\varsigma_\zs{t}=(y_\zs{t},v_\zs{t}X_\zs{t})$, where $(y_\zs{t})_\zs{0\le t\le T}$
is a deterministic function taking values  in $\bbr^d$
 such that
\begin{equation}\label{2.6-1}
\|y\|_T^2 = \int^T_\zs{0}\,|y_\zs{t}|^2\d t\,<\,\infty
\,.
\end{equation}

The process $(X_\zs{t})_\zs{0\le t\le T}$ is defined by
equation \eqref{2.4}, which in this case has the following form
(to emphasize that the wealth process corresponds to
some control process $\varsigma$ we write $X^{\varsigma}$)
\begin{equation}\label{2.7}
\d X^{\varsigma}_\zs{t}\,=\, X^{\varsigma}_\zs{t}\,(r_\zs{t}\,-\,v_\zs{t}\,+\,y'_\zs{t}\,\theta_\zs{t})
\,\d t\,+\,X^{\varsigma}_\zs{t}\,y'_\zs{t}\,
\d W_\zs{t}\,, \quad t>0\,,\quad X^{\varsigma}_\zs{0}\,=\,x\,.
\end{equation}
We denote by $\cU$
the set of all such control processes $\varsigma$.

Note that for every $\varsigma\in\cU$, by It\^o's formula,
equation \eqref{2.7} has solution
\begin{equation}\label{2.8}
 X^{\varsigma}_\zs{t}\,=\,x\,e^{R_t-V_t+(y,\theta)_t}\,\cE_\zs{t}(y)\,,
\end{equation}
where
\begin{equation}\label{2.9}
R_t=\int^t_0 r_\zs{u}\d u\,,\quad V_t=\int^t_0 v_\zs{u}\d u\quad\mbox{and}\quad
(y,\theta)_t=\int^t_0 y'_\zs{u}\,\theta_\zs{u}\d u\,.
\end{equation}
Moreover,  $\cE(y)$ denotes the stochastic exponential defined as
$$
\cE_\zs{t}(y)=
\exp\Big(\int^t_0 y'_\zs{u}\d W_\zs{u}
-\frac{1}{2} \int^t_0 |y_u|^2\d u\Big)\quad t\ge 0\,.
$$
Therefore, for every $\varsigma\in\cU$
the process $(X^{\varsigma}_\zs{t})_\zs{0\le t\le T}$ is positive and continuous.

We consider $\cU$  as a first class of control processes for equation \eqref{2.4}, for which
we can solve the control problem explicitly and interpret
its solution.
This is due to the fact, as we shall see in Section~\ref{subsec:2.4},
that because of the Gaussianity of the log-process we have explicit representations of the risk measures.

It is clear that
the  behaviour of investors
 in the model \eqref{2.1}
depends on the coefficients
$(r_\zs{t})_\zs{0\le t\le T}$, $(\mu_\zs{t})_\zs{0\le t\le T}$ and
$(\sigma_\zs{t})_\zs{0\le t\le T}$ which in our case are
nonrandom known functions and as we will see below (Corollary~\ref{Co.3.3})
for the "equlibrate utility functions" case
optimal strategies are deterministic, i.e. belong to this class.

A natural generalisation of $\cU$ is the following set of controls.

\begin{definition}\label{De.2.1}
Let $T>0$ be a fixed investment horizon.
A stochastic control process
$\varsigma=(\varsigma_\zs{t})_\zs{0\le t\le T}=((y_\zs{t},c_\zs{t}))_\zs{0\le t\le T}$ is called
{\em admissible} if it is $(\cF_\zs{t})_\zs{0\le t\le T}$-progressively measurable with values in $\bbr^d\times [0,\infty)$,
and equation \eqref{2.4} has a unique strong a.s. positive continuous solution
$(X^{\varsigma}_\zs{t})_\zs{0\le t\le T}$ on $[0\,,\,T]$.
We denote by $\cV$ the class of all {\em admissible control processes}.
\end{definition}

\subsection{The Cost Functions}
\label{subsec:2.3}

We investigate different cost functions, each leading
to a different optimal control problem.
We assume that the investor wants to optimize expected utility of consumption over the time interval $[0,T]$ and wealth $X^\varsigma_T$ at the end of the investment horizon.
For initial endowment $x>0$ and a control process $(\varsigma_t)_\zs{0\le t\le T}$ in $\cV$,
we introduce the {\em cost function}
$$
J(x,\varsigma):=\E_x\,\left(\int^T_0\,U(c_t)\,\d t\,+\,h(X^{\varsigma}_\zs{T})\right)\,,
$$
where $U$ and $h$ are {\em utility functions}.
This is a classical approach to the problem; see Karatzas and Shreve~\cite{KaSh2}, Chapter~6.

Here $\E_\zs{x}$ is the expectation operator conditional on  $X^{\varsigma}_\zs{0}=x$. For both utility functions we choose
 $U(z)=z^{\gamma_\zs{1}}$ and $h(z)=z^{\gamma_\zs{2}}$ for $z\ge 0$ with
$0<\gamma_\zs{1},\gamma_\zs{2}\le 1$,
corresponding to the cost function
\begin{equation}\label{2.10}
J(x,\varsigma):=\E_x\,\left(\int^T_0\,c_t^{\gamma_\zs{1}}\,\d t\,+\,(X^{\varsigma}_\zs{T})^{\gamma_\zs{2}}\right)\,.
\end{equation}
For $\gamma<1$ the utility function $U(z)=z^\gamma$ is concave and is called a power (or HARA) utility function.
We include the case of $\gamma=1$, which corresponds to simply optimizing expected consumption and terminal wealth. In combination with a downside risk bound this allows us in principle to dispense
with the utility function, where in practise one has to choose the parameter $\gamma$.
In the context of this paper it also allows us to separate the effect of the utility function and the risk limit.

\subsection{The Downside Risk Measures}\label{subsec:2.4}

As risk measures we use modifications of the Value-at-Risk and the Expected Shortfall as introduced in Emmer, Kl\"uppelberg and Korn~\cite{EmKlKo}. They can be summarized under the notion of Capital-at-Risk and limit the possibility of excess losses over the riskless investment.
In this sense they reflect a capital reserve.
If the resulting risk measure is negative (which can happen in certain situations) we interpret this as an additional possibility for investment.
For further interpretations we refer to~\cite{EmKlKo}.

To avoid non-relevant cases we consider only $0<\alpha<1/2$.

\begin{definition}\label{def2.2} {\em [Value-at-Risk (VaR)]}\\
Define for initial endowment $x>0$, a control process $\varsigma\in\cU$ and $0<\alpha\le 1/2$ the {\em Value-at-Risk (VaR)} by
$$
{\rm VaR}_t(x,\varsigma,\alpha):=x\,e^{R_t}-\lambda_t\,,
\quad t\ge 0\,,
$$
where $\lambda_\zs{t}=\lambda_t(x,\varsigma,\alpha)$ is the
$\alpha$-quantile
of $X^{\varsigma}_\zs{t}$, i.e.
$$
\lambda_t=\inf\{\lambda\ge 0\,:\,
\P(X^{\varsigma}_\zs{t}\,\le\,\lambda)\ge \alpha\}\,.
$$
\end{definition}

\begin{corollary}\label{Co.2.3}
In the situation of Definition~\ref{def2.2},
for every $\varsigma\in\cU$  the $\alpha$-quantile  $\lambda_t$ is given by
$$
\lambda_t\,=\,x\,
\exp\left(R_t-V_t+(y,\theta)_t-\frac{1}{2}\|y\|^2_\zs{t}
-|z_\alpha|\|y\|_\zs{t}\right)\,, \quad t\ge0\,,
$$
where $z_\alpha$ is the $\alpha$-quantile of the standard normal distribution,
and the other quantities are defined in \eqref{2.3} and \eqref{2.9}.
\end{corollary}
\bigskip

We define the {\em level risk function} for some coefficient $0<\zeta<1$ as
\begin{equation}\label{2.11}
\zeta_t(x)\,=\,\zeta\,x\,e^{R_t}\,,\quad t\in [0,T]\,.
\end{equation}
We consider only controls $\varsigma\in\cU$ for which the
Value-at-Risk is bounded by the level function
(\ref{2.11}) over the interval $[0,T]$; i.e. we require
\begin{equation}\label{2.12}
\sup_\zs{0\le t\le T}\,\frac{{\rm VaR}_t(x,\varsigma,\alpha)}{\zeta_t(x)}\,\le\,1\,.
\end{equation}

We have formulated the time-dependent risk bound
in the same spirit as we have defined the risk measures, which are based on a comparisn of the minimal possible wealth in terms of a low quantile to the pure bond investment. The risk bound now limits the admissible risky strategies to those, whose risk compared to the pure bond portfolio, represented by $\zeta$, remains uniformly bounded over the investment interval.

Our next risk measure is an analogous modification of the  {\em Expected Shortfall} (ES).

\begin{definition}\label{def2.4} {\em [Expected Shortfall (ES)]}\\
Define for initial endowment $x>0$, a control process $\varsigma\in\cU$ and $0<\alpha\le 1/2$
\begin{align*}
m_\zs{t}(x,\varsigma,\alpha)&\,=\,
\E_\zs{x} (X^\varsigma_\zs{t}\,|\,X^\varsigma_\zs{t}\le\,\lambda_\zs{t})\,,\quad t\ge0\,,
\end{align*}
where $\lambda_t(x,\varsigma,\alpha)$ is the $\alpha$-quantile
of $X^{\varsigma}_\zs{t}$.
The {\em Expected Shortfall (ES)} is then defined as
$$
{\rm ES}_\zs{t}(x,\varsigma,\alpha)
\,=\,xe^{R_t}\,-\,m_\zs{t}(x,\varsigma,\alpha)\,,\quad t\ge0\,.
$$
\end{definition}

The following result is an analogon of Corollary~\ref{Co.2.3}.

\begin{corollary}\label{Co.2.5}
In the situation of Definition~\ref{def2.4},
for any $\varsigma\in\cU$
the quantity $m_t=m_\zs{t}(x,\varsigma,\alpha)$ is given by
\begin{align*}
m_\zs{t}(x,\varsigma,\alpha) &\,=\,x\,
F_\zs{\alpha}\,(|z_\zs{\alpha}|+\|y\|_\zs{t})\,
e^{R_t+(y,\theta)_t-V_t}\,,\quad t\ge 0\,,
\end{align*}
where where $z_\alpha$ is the $\alpha$-quantile of the standard normal distribution and
$$
F_\zs{\alpha}(z)\,=\,
\frac{\int^{\infty}_\zs{z}\,e^{-t^2/2}\,\d t} {\int^{\infty}_\zs{|z_\zs{\alpha}|}\,e^{-t^2/2}\,\d t}
\,,\quad z\ge 0\,.
$$
\end{corollary}

We shall consider all controls $\varsigma\in\cU$, for which the Expected Shortfall
 is bounded by the level function (\ref{2.11}) over the interval
$[0,T]$, i.e. we require
\begin{equation}\label{2.13}
\sup_\zs{0\,\le\,t\,\le\,T}\,
\frac{{\rm ES}_\zs{t}(x,\varsigma,\alpha)}
{\zeta_\zs{t}(x)}\,\le\,1\,.
\end{equation}

\begin{remark}\rm
\label{Re.2.6}
(i) \, The coefficient $\zeta$ introduces some risk aversion behaviour into the model.
In that sense it acts similarly as a utility function does.
The difference, however, is that $\zeta$ has a clear interpretation,
and every investor can choose and understand the influence of $\zeta$ with respect to the corresponding risk measures.\\[2mm]
(ii) \, If $\|y\|_t=0$ for all $t\in[0,T]$, then ${\rm VaR}_t(x,\varsigma,\alpha)={\rm ES}_\zs{t}(x,\varsigma,\alpha)=xe^{R_t}(1-e^{-V_t})$, $0\le t\le T$.
On the other hand, if  $\|y\|_t>0$ for $t\in[0,T]$, then
$$\lim_{\alpha\to 0}{\rm VaR}_t(x,\varsigma,\alpha)=
\lim_{\alpha\to 0}{\rm ES}_\zs{t}(x,\varsigma,\alpha)= xe^{R_t}\,.$$
This means that the choice of $\alpha$ influences the risk bounds \eqref{2.12} and \eqref{2.13}.
Note, however, that $\alpha$ is chosen by the regulatory authorities, not by the investor.
The investor only chooses the value $\zeta$. If $\zeta$ is near 0 the risk level is rather low,
whereas for $\zeta$ close to 1 the risk level is rather high, indeed in such case the risk bounds may not be restrictive at all.
\end{remark}

\section{Problems and Solutions}
\label{sec:3}
\setcounter{equation}{0}

In the situation of Section~\ref{sec:2} we are interested in the solutions to different optimization problems.
Throughout we assume a fixed investment horizon $T>0$.

In the following we first present the solution to the unconstrained problem and then study the constrained problems.
The constraints are in terms of
risk bounds with respect to downfall risks like VaR and ES
defined by means of a quantile.

\subsection{The Unconstrained Problem }
\label{subsec:3.1}

We consider two regimes with cost functions \eqref{2.10} for
$0<\gamma_\zs{1}, \gamma_\zs{2}<1$  and for $\gamma_\zs{1}=\gamma_\zs{2}=1$.
We include the case of $\gamma_\zs{1}=\gamma_\zs{2}=1$ for further referencing, although it makes
economically not much sense without a risk constraint.
The mathematical  treatment of the two cases is completely different by nature.

\newpage
\begin{problem}\label{P}
\end{problem}
\begin{equation*}
\max_\zs{\varsigma\in\cV}\,J(x,\varsigma)\,.
\end{equation*}

\begin{theorem}\label{Th.3.1}
Consider Problem~\ref{P} with $\gamma_\zs{1}= \gamma_\zs{2}=1$.
Assume a riskless interest rate
$r_\zs{t}\ge 0$ for all $t\in [0,T]$.\\
If $\|\theta\|_\zs{T}>0$, then
$$
\max_\zs{\varsigma\in\cU}\,J(x,\varsigma)\,=\,\infty\,.
$$
If $\|\theta\|_\zs{T}=0$, then a solution exists and the optimal value of
 $J(x,\varsigma)$ is
given by
$$
\max_\zs{\varsigma\in\cU}\,J(x,\varsigma)\,=\,J(x,\varsigma^*)\,=\,x\,e^{R_\zs{T}}\,,
$$
corresponding  to the optimal control $\varsigma_t^*=(y_t^*,0)$ for all $0\le t\le T$  with arbitrary deterministic
square integrable function $(y^*_\zs{t})_\zs{0\le t\le T}$.
In this case   the optimal wealth process
$(X^*_t)_{0\le t\le T}$ satisfies the following equation
\begin{equation}\label{3.1}
\d X^*_\zs{t}=X^*_\zs{t} r_\zs{t}\d t\,+\,X^*_\zs{t}\,
(y^*_{\zs{t}})'\,
\d W_\zs{t}\,,\quad X^*_\zs{0}\,=\,x\,.
\end{equation}
\end{theorem}


Consider now Problem~\ref{P} for $0<\gamma_1,\gamma_2<1$.
To formulate the solution we define functions
\begin{align}\label{3.2}
A_\zs{1}(t)=\gamma_\zs{1}^{q_1}
\int^T_\zs{t}\,e^{\int^s_\zs{t}\beta_\zs{1}(u)\d u}\,\d s
\quad\mbox{and}\quad
A_\zs{2}(t)=\gamma_\zs{2}^{q_2}\,e^{\int^T_\zs{t}\beta_\zs{2}(u)\d u}\,,\quad 0\le t\le T\,,
\end{align}
where $q_i=(1-\gamma_i)^{-1}$ and
$\beta_\zs{i}(t)=(q_i-1)( r_\zs{t}+\frac{q_i}{2}|\theta_\zs{t}|^2)$.
Moreover, for all $0\le t\le T$ and $x>0$ we define the function
$g(t,x)>0$ as solution to
\begin{align}\label{3.3}
A_\zs{1}(t)\,g^{-q_1}(t,x)+A_\zs{2}(t)\,g^{-q_2}(t,x)=x
\end{align}
and
$$
p(t,x)=q_1 A_\zs{1}(t)\,g^{-q_1}(t,x)+q_2 A_\zs{2}(t)\,g^{-q_2}(t,x)\,.
$$

\begin{theorem}\label{Th.3.2}
Consider Problem~\ref{P} for $0<\gamma_\zs{1}, \gamma_\zs{2}<1$.
The optimal value of $J(x,\varsigma)$ is given by
$$
\max_\zs{\varsigma\in\cV}\,J(x,\varsigma)\,=\,J(x,\varsigma^*)\,
=\frac{A_\zs{1}(0)}{\gamma_\zs{1}}g^{1-q_1}(0,x)+
\frac{A_\zs{2}(0)}{\gamma_\zs{2}}g^{1-q_2}(0,x)\,,
$$
where the optimal control $\varsigma^*=(y^*,c^*)$ is for all $0\le t\le T$ of the form
\begin{equation}\label{3.4}
\left\{
\begin{array}{rl}
y^*_\zs{t}&=\dfrac{p(t,X^*_\zs{t})}{X^*_\zs{t}}\,\theta_\zs{t}
\quad\left( \pi^*_\zs{t}=\dfrac{p(t,X^*_\zs{t})}{X^*_\zs{t}}\,(\si_t\si_t')^{-1}(\mu_t-r_t\bf 1 )\right)\,;\\[6mm]
c^*_\zs{t}&=\left(\dfrac{\gamma_\zs{1}}{g(t,X^*_\zs{t})}
\right)^{q_1}\,.
\end{array}
\right.
\end{equation}
The optimal wealth process $(X^*_t)_{0\le t\le T}$ is the solution to
\begin{equation}\label{3.5}
\d X^*_\zs{t}\,=a^*(t,X^*_\zs{t})\d t+(b^*(t,X^*_\zs{t}))'\d W_\zs{t}\,, \quad
\  X^*_\zs{0}\,=\,x\,,
\end{equation}
where
$$
a^*(t,x)=r_\zs{t}x+p(t,x)\,|\theta_\zs{t}|^2
-\left(\frac{\gamma_\zs{1}}{g(t,x)}\right)^{q_1}
\quad\mbox{and}\quad
b^*(t,x)=p(t,x)\,\theta_\zs{t}\,.
$$
\end{theorem}

The following result can be found Example~6.7 on p.~106 in Karatzas and Shreve
\cite{KaSh2}; its proof here is based on the martingale method.

\begin{corollary}\label{Co.3.3}
Consider Problem~\ref{P} for $\gamma_\zs{1}=\gamma_\zs{2}=\gamma\in(0,1)$ and define
\begin{equation}\label{3.6}
\wt{g}_\zs{\gamma}(t)=\exp\left(\gamma R_\zs{t}+\frac{q-1}{2}\,\|\theta\|^2_\zs{t}\right)\quad \mbox{and}\quad q=\frac{1}{1-\gamma}\,.
\end{equation}
Then the optimal value of $J(x,\varsigma)$  is given by
$$
J^*(x)\,=\,\max_\zs{\varsigma\in\cV}\,J(x,\varsigma)\,=\,J(x,\varsigma^*)\,
=x^{\gamma}\,\left(\|\wt{g}_\zs{\gamma}\|_\zs{q,T}^{q}+\wt{g}_\zs{\gamma}^{q}(T)\right)^{1/q}\,,
$$
where the optimal control $\varsigma^*=(y^*,c^*)$ is for all $0\le t\le T$ of the form
\begin{equation}\label{3.7}
\left\{
\begin{array}{rl}
y^*_\zs{t}&=\dfrac{\theta_\zs{t}}{1-\gamma} \quad
\left(
\pi^*_\zs{t}=\dfrac{(\si_t\si_t')^{-1}(\mu_t-r_t{\bf 1} )}{1-\gamma}
\right)\,;\\[6mm]
c^*_\zs{t}&=v^*_\zs{t}X^*_t
\quad\mbox{and}\quad
v^*_\zs{t}=\dfrac{\wt{g}_\zs{\gamma}^{q}(t)}{\wt{g}_\zs{\gamma}^{q}(T)+\int^T_t\,\wt{g}_\zs{\gamma}^{q}(s)\,\d s}\,.
\end{array}
\right.
\end{equation}
The optimal wealth process $(X^*_t)_{0\le t\le T}$ is given by
\begin{equation}\label{3.8}
\d X^*_\zs{t}\,=\,X^*_\zs{t}\left(r_\zs{t}\,-\,v^*_\zs{t}\,+
\,\frac{|\theta_\zs{t}|^2}{1-\gamma}\right)\d t
+X^*_\zs{t}\frac{\theta'_\zs{t}}{1-\gamma}\,\d W_\zs{t}\,, \quad
  X^*_\zs{0}=x\,.
\end{equation}
\end{corollary}

\begin{remark}\label{Re.3.4}\rm
 Note that Problem~\ref{P} for different
$0<\gamma_\zs{1}<1$ and $0<\gamma_\zs{2}<1$
was also investigated by Karatzas and Shreve~\cite{KaSh2}.
For  H\"older continuous  market coefficients they find by
the martingale method  an implicit ``feedback form''  of the optimal solution in their Theorem~8.8.
In contrast, Theorem~\ref{Th.3.2} above gives the optimal solution in
``explicit feedback form'' for quite general market coefficients.
Our proof is based on a special
version of a verification theorem for stochastic optimal control
problems, which allows for c\`adl\`ag coefficients.
\end{remark}

\subsection{Value-at-Risk as Risk Measure}
\label{subsec:3.2}

For the Value-at-Risk we consider again the cost function \eqref{2.10} and, as before,
we consider different regimes for $0<\ga_\zs{1}, \ga_\zs{2}< 1$ and $\ga_\zs{1},\ga_\zs{2}= 1$.

\bigskip
\begin{problem}\label{P1}
\end{problem}
$$
\max_\zs{\varsigma\in\cU}\,J(x,\varsigma)\quad
\mbox{subject to}\quad\sup_\zs{0\le t\le T}\,
\frac{{\rm VaR}_t(x,\varsigma,\alpha)}{\zeta_t(x)}\,\le\,1\,.
$$

\bigskip

To formulate the solution let $z_\alpha$ be the normal $\alpha$-quantile for $0<\al\le 1/2$ and
the constant  $\zeta\in (0,1)$ as in \eqref{2.11}.
Obviously, for $\alpha\to 0$ we have $|z_\alpha|\to\infty$ and, hence, the quotient
in \eqref{2.12} tends to $1/\zeta>1$.
This means that the bound can be restrictive.
We define for $\theta$ as in \eqref{2.3} the following quantity
\begin{equation}\label{3.9}
\rho^*_\zs{\rm VaR}=\sqrt{\left(|z_\alpha|-\|\theta\|_\zs{T}\right)^2-2\ln(1-\zeta)}
-(|z_\alpha|-\|\theta\|_\zs{T})\,.
\end{equation}

\begin{theorem}\label{Th.3.5}
Consider Problem~\ref{P1} for $\gamma_\zs{1}=\gamma_\zs{2}=1$.
Assume a riskless interest rate $r_\zs{t}\ge 0$  for all $t\in [0,T]$.
Then for
\begin{equation}\label{3.10}
\max(0,1-e^{z^2_\zs{\alpha}/2-|z_\zs{\alpha}|\|\theta\|_\zs{T}})< \zeta<1
\end{equation}
the optimal value of $J(x,\varsigma)$ is given by
\begin{equation}\label{3.11}
\max_\zs{\varsigma\in\cU}\,J(x,\varsigma) =J(x,\varsigma^*)\,=x\,e^{\rho^*_\zs{\rm VaR}\|\theta\|_\zs{T}+R_\zs{T}}\,.
\end{equation}
If $\|\theta\|_\zs{T}>0$, then the optimal control $\varsigma^*=(y^*,v^*X^*)$ is for all $0\le t\le T$ of the form
\begin{equation}\label{3.12}
y^*_\zs{t}\,=\,\rho^*_\zs{\rm VaR}\,\frac{\theta_t}{\|\theta\|_\zs{T}}
\quad\quad
\Big(
\pi^*_\zs{t}=
\rho^*_\zs{\rm VaR}\frac{(\sigma_\zs{t}\sigma'_\zs{t})^{-1}}{\|\theta\|_\zs{T}} (\mu_t-r_t{\bf 1})
\Big)
\quad\mbox{and}\quad
v^*_\zs{t}=0\,.
\end{equation}
The optimal wealth process $(X^*_\zs{t})_{0\le t\le T}$ is given by
$$
\d X^*_\zs{t}\,=\,X^*_\zs{t}\, \Big(r_\zs{t}+
\rho^*_\zs{\rm VaR}\,\frac{|\theta_t|^2}{\|\theta\|_\zs{T}}\Big)\,\d t
+\,X^*_\zs{t}\,\rho^*_\zs{\rm VaR}\,\frac{\theta'_t}{\|\theta\|_\zs{T}}\,
\d W_\zs{t}\,, \quad  X^*_\zs{0}\,=\,x\,.
$$
If $\|\theta\|_\zs{T}=0$, then the optimal value of $J(x,\varsigma)$ is given by
\begin{equation}\label{3.13}
\max_\zs{\varsigma\in\cU}\,J(x,\varsigma) =J(x,\varsigma^*)\,=x\,e^{R_\zs{T}}\,,
\end{equation}
corresponding to the optimal control $\varsigma^*_\zs{t}=(y^*_\zs{t}\,,0)$ for $0\le t\le T$
with  arbitrary deterministic function $(y^*_\zs{t})_\zs{0\le t\le T}$ such that
$$\|y^*\|_\zs{T}\le \rho^*_\zs{\rm VaR}
=\sqrt{z^2_\alpha-2\ln(1-\zeta)}-|z_\alpha|\,.
$$
In this case the optimal wealth process
$(X_t^*)_\zs{0\le t\le T}$ satisfies equation \eqref{3.1}.
\end{theorem}

\begin{remark}\label{Re.3.6}\rm
(i) \, For $|z_\zs{\alpha}|\ge 2\|\theta\|_\zs{T}$ condition \eqref{3.10} gives a lower bound 0; i.e.\\ $0<\zeta<1$.
If $|z_\zs{\alpha}|< 2\|\theta\|_\zs{T}$, then condition \eqref{3.10} translates to
$$
1-e^{z^2_\zs{\alpha}/2-|z_\zs{\alpha}|\|\theta\|_\zs{T}}< \zeta<1\,;
$$
i.e. we obtain a positive lower bound.\\[2mm]
(ii) \, The optimal strategy implies that there will be no consumption throughout the investment horizon.
This is due to the fact that the wealth we expect by investment
is so attractive that we continue to invest everything.
Note that the solution is the same as the solution to the problem without possible
consumption.
\end{remark}

Now we present a sufficient condition for which
the optimal unconstrained strategy \eqref{3.7}--\eqref{3.8}
is solution for Problem~\ref{P1} in the case
$\gamma_\zs{1}=\gamma_\zs{2}=\gamma\in (0,1)$. For this we introduce
the following functions:
$$
\wt\kappa(\gamma)=\frac{\|\wt{g}_\zs{\gamma}\|_\zs{q,T}^{q} }{\|\wt{g}_\zs{\gamma}\|_\zs{q,T}^{q}+\wt{g}_\zs{\gamma}^{q}(T)}
=1- e^{-V^*_T}= 1- e^{-\int^{T}_\zs{0}\,v^{*}_\zs{t}\d t}\,,
$$
where $(v^{*}_\zs{t})_\zs{0\le t\le T}$ is the optimal
 consumption rate introduced in
\eqref{3.7}. By setting $\wt{l}(\gamma)=\ln\left(1-\wt\kappa(\gamma)\right)$
we define
$$
l_*(\gamma)=
\left\{\begin{array}{ll}
-q\|\theta\|_\zs{T}\,|z_\alpha|+\wt{l}(\gamma) \quad & \mbox{for} \ \  0<\gamma\le 1/2\,; \\[5mm]
-q\|\theta\|_\zs{T}\,|z_\alpha|+\wt{l}(\gamma)-\frac{q(q-2)}{2}\|\theta\|^2_\zs{T}
\quad &  \mbox{for}\ \  1/2<\gamma< 1\,.
\end{array}\right.
$$

\begin{theorem}\label{Th.3.9}
Consider Problem~\ref{P1} with $\gamma_\zs{1}=\gamma_\zs{2}=\gamma\in (0,1)$.
Assume a riskless interest rate $r_\zs{t}\ge 0$  for all $t\in [0,T]$ and
\begin{equation}\label{3.21}
1-e^{l_*(\gamma)}\,\le\,\zeta<1\,.
\end{equation}
Then the optimal solution
is given by \eqref{3.7}--\eqref{3.8}; i.e. it is equal
 to the solution of the unconstrained problem.
\end{theorem}

\begin{remark}\label{Re.3.10}\rm
Theorem~\ref{Th.3.9}  does not hold for $\gamma_\zs{1} \neq\gamma_\zs{2}$,
since the solution \eqref{3.4}  does not belong to $\cU$.
\end{remark}

To formulate the result for different $\gamma_i$ ($i=1,2$) we introduce the following function
  for $0\le\kappa\le 1$
\begin{equation}\label{3.22}
G(x,\kappa):=x^{\gamma_\zs{1}}\kappa^{\gamma_\zs{1}}\|\wh{g}_\zs{1}\|_\zs{q,T}+  x^{\gamma_\zs{2}}(1-\kappa)^{\gamma_\zs{2}} \wh{g}_\zs{2}(T)\,,\quad  x>0\,,
\end{equation}
where $q=(1-\gamma_\zs{1})^{-1}$, $\wh{g}_\zs{i}= \wh{g}_\zs{\gamma_\zs{i}}$ and
$$
\wh{g}_\zs{\gamma}=e^{\gamma R_\zs{t}}=e^{\gamma \int^t_\zs{0}r_\zs{u}\d u}\,.
$$
Moreover, for $x>0$ we set
\begin{equation}\label{3.23}
\kappa_\zs{*}(x) =\arg\max_\zs{0\le \kappa\le 1}G(x,\kappa)\,.
\end{equation}
Note that for $0<\gamma_\zs{1}<1$ and  $0<\gamma_\zs{2}\le 1$
this function is strictly positive for all $x>0$; i.e. $0<\kappa_\zs{*}(x)\le 1$.
It is easy to see that in the case $\gamma_\zs{1}=\gamma_\zs{2}=:\gamma$ the function
$\kappa_*(x)$ is independent of $x$ and equals to

\begin{equation}\label{3.20}
\wh\kappa(\gamma)=\frac{\|\wh{g}_\zs{\gamma}\|^{q}_\zs{q,T}}
{\|\wh{g}_\zs{\gamma}\|^{q}_\zs{q,T}+
\wh{g}_\zs{\gamma}^q(T)}\,.
\end{equation}

\begin{theorem}\label{Th.3.12}
Consider Problem~\ref{P1} with $0<\gamma_\zs{1}<1$ and $0<\gamma_\zs{2}\le 1$.
Assume a riskless interest rate $r_\zs{t}\ge 0$  for all $t\in [0,T]$ and
\begin{equation}\label{3.24}
0<\zeta<\min\{\kappa_*(x)\,,\,\wh\kappa(\gamma_\zs{1})\}\,.
\end{equation}
Moreover, assume that
\begin{equation}\label{3.25}
|z_\zs{\alpha}|\ge \left(1+
\frac{\max\{\gamma_\zs{1}\,,\,\gamma_\zs{2}\}}{1-\zeta}\,
\frac{1}{\frac{\partial}{\partial\zeta}\ln\,G(x,\zeta)}\right)\|\theta\|_\zs{T}\,.
\end{equation}
Then the optimal value of $J(x,\varsigma)$  is given by
\begin{equation}\label{3.26}
\max_\zs{\varsigma\in\cU}\,J(x,\varsigma)=\,J(x,\varsigma^*)=
x^{\gamma_\zs{1}}\zeta^{\gamma_\zs{1}}\|\wh{g}_\zs{1}\|_\zs{q,T}+
x^{\gamma_\zs{2}}(1-\zeta)^{\gamma_\zs{2}} \wh{g}_\zs{2}(T)\,,
\end{equation}
where the optimal control $\varsigma^*=(y^*,v^*X^*)$ is for all $0\le t\le T$ of the form
\begin{equation}\label{3.27}
y^*_\zs{t}=0\quad (\pi^*_\zs{t}=0) \ \
\mbox{and} \ \
v^*_\zs{t}=\frac{\zeta\, \wh{g}^q_\zs{1}(t)}
{\|\wh{g}_\zs{1}\|^{q}_\zs{q,T}-\zeta\, \|\wh{g}_\zs{1}\|^{q}_\zs{q,t}}\,.
\end{equation}
The optimal wealth process $(X^*_t)_{0\le t\le T}$ is given by the
deterministic function
\begin{equation}\label{3.28}
X^*_\zs{t}=x\,e^{R_t}\,
\frac{\|\wh{g}_\zs{1}\|^{q}_\zs{q,T}-\zeta \|\wh{g}_\zs{1}\|^{q}_\zs{q,t}}
{\|\wh{g}_\zs{1}\|^{q}_\zs{q,T}}
=x\,\frac{\zeta}{v^*_\zs{t}}\,e^{R_t}\,,\quad 0\le t\le T\,.
\end{equation}
\end{theorem}


\begin{remark}\label{Re.3.14}\rm
 We compare now conditions
\eqref{3.24}--\eqref{3.25} for $\gamma_\zs{1}=\gamma_\zs{2}=\gamma\in (0,1)$
with condition \eqref{3.21}.
 Making use of
the notation in \eqref{3.6} we obtain
$$
\wt{g}_\zs{\gamma}(t)= \wh{g}_\zs{\gamma}(t) e^{\frac{q-1}{2}\|\theta\|^2_\zs{t}}\ge \wh{g}_\zs{\gamma}(t)\,.
$$
Taking this inequality into account we find that in the case
$0<\gamma\le 1/2$\\ (i.e. $1<q\le 2$),  the function $e^{l_*(\gamma)}$ is bounded above by
\begin{align*}
e^{l_*(\gamma)}=\frac{\wt{g}_\zs{\gamma}^{q}(T)e^{-q\|\theta\|_\zs{T}\,|z_\alpha|}}{\|\wt{g}_\zs{\gamma}\|_\zs{q,T}^{q}+
\wt{g}_\zs{\gamma}^{q}(T)}
\le\frac{\wh{g}_\zs{\gamma}^{q}(T)e^{-q(\|\theta\|_\zs{T}\,|z_\alpha|-\frac{q-1}{2}\|\theta\|^2_\zs{T})}}
{\|\wh{g}\|_\zs{q,T}^{q}+\wh{g}_\zs{\gamma}^{q}(T)}\,.
\end{align*}
Moreover, condition \eqref{3.25} implies
$|z_\alpha|\ge \|\theta\|_\zs{T}$. Therefore, taking into account that
 $1<q\le 2$  we obtain
$$
e^{-q(\|\theta\|_\zs{T}\,|z_\alpha|-\frac{q-1}{2}\|\theta\|^2_\zs{T})}
\le 1\,.
$$
Hence,
$$
e^{l_*(\gamma)}
\le\frac{\wh{g}_\zs{\gamma}^{q}(T)}{\|\wh{g}\|_\zs{q,T}^{q}+\wh{g}_\zs{\gamma}^{q}(T)}=1-\wh\kappa(\gamma)\,.
$$
Similarly, for $1/2<\gamma<1$ (i.e. $q>2$),
\begin{align*}
e^{l_*(\gamma)}
\le\frac{\wh{g}_\zs{\gamma}^{q}(T)e^{-\frac{q}{2}\|\theta\|^2_\zs{T}}}
{\|\wh{g}\|_\zs{q,T}^{q}+\wh{g}_\zs{\gamma}^{q}(T)}\le 1- \wh\kappa(\gamma)\,.
\end{align*}
So we have shown that
$1-e^{l_*(\gamma)}\ge \wh\kappa(\gamma)$, i.e.
condition \eqref{3.21} is complementary to conditions \eqref{3.24}-\eqref{3.25}. 
\end{remark}

We present an example for further illustration.

\begin{example}\label{Re.3.13}\rm
To clarify conditions \eqref{3.24}--\eqref{3.25}  consider again $\gamma_\zs{1}=\gamma_\zs{2}=\gamma\in (0,1)$ and $r_\zs{t}\equiv r>0$.
We shall investigate what happens for $T\to\infty$.
First we calculate
$$
\kappa_*(x)
=
\wh\kappa(\gamma)=
\frac{\int^T_\zs{0}\,e^{q\gamma r t}\,\d t}{\int^T_\zs{0}\,e^{q\gamma r t}\,\d t+e^{q\gamma r T}}
=
\frac{1-e^{-q\gamma r T}}{ 1+q\gamma r-e^{-q\gamma r T}}\sim
\frac{1}{ 1+q\gamma r}
$$
as $T\to\infty$, where $q=(1-\gamma)^{-1}$. Thus,
condition \eqref{3.24} yields for $T\to \infty$ approximately
$$
0<\zeta<\frac{1}{ 1+q\gamma r}\,.
$$
The function \eqref{3.22} has the following form
$$
G(x,\kappa)=x^{\gamma}
e^{\gamma r T}
\left(\kappa^\gamma A(T)
+
(1-\kappa)^\gamma
\right)
\quad
\mbox{with}
\quad
A(T)=\left(
\int^T_\zs{0}\,e^{-q\gamma r t}\,\d t
\right)^{1/q}\,.
$$
For the partial derivative with respect to $\zeta$ we calculate
$$
\frac{\partial}{\partial\zeta}\ln\,G(x,\zeta)
=\gamma \frac{\zeta^{\gamma-1}A(T)-(1-\zeta)^{\gamma-1}}{\zeta^{\gamma}A(T)+(1-\zeta)^{\gamma}}.
$$
Since
\beao
\frac{\max\{\gamma_\zs{1}\,,\,\gamma_\zs{2}\}}{1-\zeta}\,
\frac{1}{\frac{\partial}{\partial\zeta}\ln\,G(x,\zeta)}
=
\frac{\zeta^{\gamma+1}A(T)+\zeta(1-\zeta)^\gamma}{\zeta^{\gamma}(1-\zeta)A(T)-\zeta(1-\zeta)^\gamma}
= \mbox{\em O}(\zeta)
\quad\mbox{as}\quad \zeta\to 0\,,
\eeao
condition \eqref{3.25} implies $|z_\zs{\alpha}| > \|\theta\|_\zs{T}$ approximately for $\zeta\to 0$.
Moreover, the optimal consumption \eqref{3.27} is given by
$$
v^*_\zs{t}=\zeta\,  \frac{ \gamma q r}
{e^{\gamma q r (T-t)}-\zeta  -(1-\zeta)e^{-\gamma q r t}}
$$
and the optimal wealth process \eqref{3.28}  is
$$
X^*_\zs{t}
=x\,\frac{e^{rt}}{\gamma q r}
\left(e^{\gamma q r (T-t)}-\zeta -(1-\zeta)e^{-\gamma q r t}\right)
\,,\quad 0\le t\le T\,.
$$
\end{example}

\begin{conclusion}\rm
The preceding results allow us to compare the optimal strategies of the unconstrained
problems and the constrained problems with VaR bound. We consider a riskless interest rate
 $r_t\ge 0$ for all $t\in[0,T]$.

When simply optimizing expectation, i.e. $\gamma_1=\gamma_2=1$,
the VaR constrain puts a limit to the investment strategy and also
 influences the optimum wealth. On the other hand,
 there is no change in the consumption,
 which is zero throughout the investment horizon
 in both cases.

For $0<\gamma_1,\gamma_2\le 1$ the optimal strategy for the utility maximization problem
 involves investment and consumption during the investment horizon;
 cf. Theorem~\ref{Th.3.5}. The influence of a VaR bound is dramatic,
 when it is valid, as it recommends the optimal strategy of no investment,
 but consumption only; cf. Theorem~\ref{Th.3.12}.
\end{conclusion}

\subsection{Expected Shortfall as Risk Measure}
\label{subsec:3.3}

The next problems concern bounds on the Expected Shortfall.

\bigskip
\begin{problem}\label{P2}
\end{problem}
\begin{equation*}
\max_\zs{\varsigma\in\cU}\,J(x,\varsigma)\quad
\mbox{subject to}\quad\sup_\zs{0\,\le\,t\,\le\,T}\,
\frac{ES_\zs{t}(x,\varsigma,\alpha)}
{\zeta_\zs{t}(x)}\,\le\,1\,.
\end{equation*}

\bigskip

\noindent
To formulate the solution for Problem~\ref{P2} we define   for $\rho\ge 0$ and
$0\le u\le 1$
\begin{equation}\label{3.29}
\psi(\rho,u)\,=\,\|\theta\|_\zs{T}\,\rho\,u^2\,+\,
\ln\,F_\zs{\alpha}\,(|z_\zs{\alpha}|\,+\,\rho\,u)\,.
\end{equation}
Moreover, we set
\begin{equation}\label{3.30}
\rho^*_{\rm ES}\,=\,\sup\{\rho>0\,:\,\psi(\rho,1)\ge\ln (1-\zeta)\}\,,
\end{equation}
where we define $\sup\{\emptyset\}=\infty$.
We formulate some properties of $\psi$ which will help us to calculate $\rho^*_{\rm ES}$.

\begin{lemma}\label{Le.3.15}
Let $0<\alpha<1/2$ such that $|z_\zs{\alpha}|\ge 2\|\theta\|_\zs{T}$.
Then $\psi$ satisfies the following properties.
\begin{enumerate}
\item[(1)]
For every $\rho>0$  the function $\psi(\rho,u)$ is strictly decreasing
for $0\le u\le 1$.
\item[(2)]
The function $\psi(\cdot,1)$ is strictly decreasing.
\item[(3)]
For every $a\le 0$ the equation $\psi(\rho,1)=a$ has a unique positive solution.\\
The equation $\psi(\rho,1)=\ln(1-\zeta)$ has solution $\rho^*_{\rm ES}$  as defined
in \eqref{3.30}. \\
For $|z_\zs{\alpha}|>1$ we have
\beam\label{rho3}
\rho^*_{\rm ES}\le
\frac{-\ln(1-z^{-2}_\zs{\alpha})-\ln(1-\zeta)}
{|z_\zs{\alpha}|-\|\theta\|_\zs{T}}\,.
\eeam
\end{enumerate}
\end{lemma}

Now we present the solution of Problem~\ref{P2},
where we start again with the situation of a small $\alpha$, where the risk bound is restrictive.

\begin{theorem}\label{Th.3.16}
Consider Problem~\ref{P2} for $\gamma_\zs{1}= \gamma_\zs{2}=1$.
Assume also that the riskless interest rate
$r_\zs{t}\ge 0$ for all $t\in [0,T]$.
 Then for every $0<\zeta<1$ and for $0<\alpha< 1/2$
such that $|z_\zs{\alpha}|\ge 2\|\theta\|_\zs{T}$
 the solution $\rho^*_{\rm ES}$ of $\psi(\rho,1)=\ln (1-\zeta)$ is finite, and
the optimal solution is given by \eqref{3.12} after
replacing $\rho^*_\zs{\rm VaR}$ by $\rho^*_\zs{\rm ES}$.
\end{theorem}

Now we consider Problem~\ref{P2} with
$\gamma_\zs{1}= \gamma_\zs{2}=\gamma\in (0,1)$.
Our next theorem concerns the case
of a loose risk bound,
 where the solution is the same as in the unconstrained case.

\begin{theorem}\label{Th.3.18}
Consider Problem~\ref{P2} for $\gamma_\zs{1}= \gamma_\zs{2}=\gamma\in (0,1)$.
Assume that the riskless interest rate
$r_\zs{t}\ge 0$ for all $t\in [0,T]$.
Assume also that $|z_\zs{\alpha}|\ge 2\|\theta\|_\zs{T}$ and
\begin{equation}\label{3.32}
1-
\left(1-\wt\kappa(\gamma)\right)e^{q\|\theta\|^2_\zs{T}}\,
F_\zs{\alpha}(|z_\zs{\alpha}|+q\|\theta\|_\zs{T})\,
\le\,\zeta\,<1\,.
\end{equation}
Then the optimal solution $\varsigma^*$ is given by \eqref{3.7}--\eqref{3.8}; i.e. it is equal to the solution of the unconstrained problem.
\end{theorem}

Now we turn to the general case of $0<\gamma_\zs{1}, \gamma_\zs{2}\le 1$, the analogon of
Theorem~\ref{Th.3.12}.

\begin{theorem}\label{Th.3.19}
Consider Problem~\ref{P2} for $0<\gamma_\zs{1}<1$ and $0<\gamma_\zs{2}\le 1$.
Assume a riskless interest rate $r_\zs{t}\ge 0$  for all $t\in [0,T]$.
Take $\kappa_\zs{*}(x)$ as in \eqref{3.23}. Assume \eqref{3.24} and
\begin{equation}\label{3.33}
|z_\zs{\alpha}|\,\ge\,
\left(2+
\frac{\max\{\gamma_\zs{1}\,,\,\gamma_\zs{2}\}}{1-\zeta}\,
\frac{1}{\frac{\partial}{\partial\zeta}\,\ln\,G(x,\zeta)}\right)\|\theta\|_\zs{T}\,.
\end{equation}
Then the optimal solution
 $\varsigma^*$ is given by \eqref{3.27}--\eqref{3.28}.
\end{theorem}

\begin{remark}\label{Re.3.20}\rm
For $|z_\zs{\alpha}|\ge 2\|\theta\|_\zs{T}$ we calculate
\begin{align*}
F_\zs{\alpha}(|z_\zs{\alpha}|+q\|\theta\|_\zs{T})&
=\frac{\int^{\infty}_\zs{|z_\zs{\alpha}|}\exp(-\frac{(t+q\|\theta\|_\zs{T})^2}{2})\d t}
{\int^{\infty}_\zs{|z_\zs{\alpha}|}\,e^{-\frac{t^2}{2}}\,\d t}
\, \le \, \exp(-2q\|\theta\|^2_\zs{T}-\frac{q^2\|\theta\|^2_\zs{T}}{2})\,.
\end{align*}
Recalling from  Remark~\ref{Re.3.14} that
$\wt{g}_\zs{\gamma}(t)= \wh{g}_\zs{\gamma}(t) e^{\frac{q-1}{2}\|\theta\|^2_\zs{t}}$
we obtain
\begin{align*}
\left(1-\wt\kappa(\gamma)\right)e^{q\|\theta\|^2_\zs{T}}\,&
F_\zs{\alpha}(|z_\zs{\alpha}|+q\|\theta\|_\zs{T})\,\le
\frac{\wt{g}_\zs{\gamma}^{q}(T)e^{-\frac{q(q+4)}{2}\|\theta\|^2_\zs{T}}}
{\|\wt{g}_\zs{\gamma}\|_\zs{q,T}^{q}+\wt{g}_\zs{\gamma}^{q}(T)}  \\
&\le \frac{\wh{g}_\zs{\gamma}^{q}(T)}{\|\wh{g}\|_\zs{q,T}^{q}+\wh{g}_\zs{\gamma}^{q}(T)}\,
e^{-\frac{5q}{2}\|\theta\|^2_\zs{T}}
\le 1-\wh\kappa(\gamma)\,,
\end{align*}
 i.e. condition \eqref{3.32} is complementary to condition \eqref{3.24}.
\halmos\end{remark}

\begin{remark}\label{Re.3.21}\rm
(i) \, It should be noted that the optimal solution \eqref{3.27}--\eqref{3.28}
for Problems~\ref{P1}
and~\ref{P2} does not depend on the coefficients
$(\mu_t)_\zs{0\le t\le T}$ and $(\sigma_t)_\zs{0\le t\le T}$ of the stock price.
These parameters only enter into
\eqref{3.24}, \eqref{3.25} and \eqref{3.33}.
Consequently, in practice it is not necessary to know these parameters precisely,
an upper bound for $\|\theta\|_\zs{T}$ suffices. \\[2mm]
(ii) \, If $\theta\equiv 0$, then conditions \eqref{3.25} and \eqref{3.33} are trivial, i.e.
the optimal solutions for  Problems~\ref{P1} and~\ref{P2}
for $0<\gamma_\zs{1}<1$ and $0<\gamma_\zs{2}\le 1$ are given
by \eqref{3.27}--\eqref{3.28} for every $0<\alpha<1/2$ and $\zeta$ satisfying \eqref{3.24}
\halmos\end{remark}

\begin{conclusion}\rm
The preceding results again allow us to compare the optimal strategies of the utility maximization problems and the constrained problems with ES bound.
The structures of the solutions are the same as for a VaR constrain, only certain values have changed.
\end{conclusion}

\section{Proofs}\label{sec:4}
\setcounter{equation}{0}

\subsection{Proof of Theorem~\ref{Th.3.1}}\label{subsec:4.1}

First we consider $\|\theta\|_\zs{T}>0$.
Define for $n\in\bbn$ the sequence of strategies $\varsigma^{(n)}=(y^{(n)},v^{(n)}X^{(n)})$ for
which $v^{(n)}=0$ and $y^{(n)}=n\theta$.
For this strategy \eqref{2.8} implies
$$
J(x,\varsigma^{(n)})= x\,e^{R_T+n\|\theta\|_T}\to\infty\quad\mbox{as}
\quad n\to\infty\,.
$$
Let now  $\|\theta\|_\zs{T}=0$. Then the cost function can be estimated above by \begin{align*}
J(x,\varsigma)&=x\,\left(\int^T_0\,e^{R_\zs{t}-V_\zs{t}}v_t\,\d t+
e^{R_\zs{T}-V_\zs{T}}\right)\\ \nonumber
&\le x\,e^{R_\zs{T}}\,\left(\int^T_0 e^{-V_\zs{t}}v_t\d t+
e^{-V_\zs{T}}\right)\\ \nonumber
&= x\,e^{R_\zs{T}}\,.
\end{align*}
Thus, every control $\varsigma$ with $v=0$ matches this upper bound.
\eproof

\subsection{Proof of Theorem~\ref{Th.3.2}}\label{subsec:4.2}

We apply the Verification Theorem~\ref{Th.A.1} to
Problem~\ref{P} for the stochastic control differential equation (\ref{2.4}).
For fixed $\vartheta\,=\,(y,c)$, where $y\in\bbr^d$ and
$c\in [0,\infty)$, the coefficients in  model (\ref{A.1}) are defined as
\begin{align*}
a(t,x,\vartheta)&\,=\,x\,(\,r_\zs{t}\,+\,y'\theta_\zs{t}\,)\,-\,c\,, \\
b(t,x,\vartheta)&\,=\,x\,|y|\,, \ \
f(t,x,\vartheta)\,=\,c^{\gamma_\zs{1}}\,,\quad h(x)\,=\,x^{\gamma_\zs{2}}\,,\quad
 0<\gamma_\zs{1},\gamma_\zs{2} <1\,.
\end{align*}
This implies immediately $\H_\zs{1}$.
Moreover, by Definition~\ref{De.2.1} the coefficients are continuous, hence
(\ref{A.2}) holds for every $\varsigma\in\cV$.\\[2mm]
To check $\H_\zs{1}-\H_\zs{3}$ we calculate
the Hamilton function \eqref{A.5} for Problem~\ref{P}. We have
$$
H(t,x,z_\zs{1},z_\zs{2})\,=\,\sup_\zs{\vartheta\in\bbr^d\times [0,\infty)}\,
H_\zs{0}(t,x,z_\zs{1},z_\zs{2},\vartheta)\,,
$$
where
\begin{align*}
H_\zs{0}(t,x,z_\zs{1},z_\zs{2},\vartheta)\,=\,
(r_\zs{t}\,+\,y'\,\theta_\zs{t})x\,z_\zs{1}\,+\,
\frac{1}{2}x^2|y|^2\,z_\zs{2}\,+\,c^{\gamma_\zs{1}}\,-\,c\,z_\zs{1}\,.
\end{align*}\noindent
For $z_\zs{2}\le 0$ we find (recall that $q_i=(1-\gamma_i)^{-1}$)
\begin{align*}
H(t,x,z_\zs{1},z_\zs{2})&\,=
\,H_\zs{0}(t,x,z_\zs{1},z_\zs{2},\vartheta_\zs{0})\\ \nonumber
&\,=\,r_\zs{t}\,x\,z_\zs{1}\, +\,\frac{1}{2|z_\zs{2}|}\,z^2_\zs{1}\,|\theta_\zs{t}|^2\,+\,\frac1{q_\zs{1}} \,\left(\frac{\gamma_\zs{1}}{z_\zs{1}}\right)^{q_\zs{1}-1}\,,
\end{align*}
where $\vartheta_\zs{0}=\vartheta_\zs{0}(t,x,z_\zs{1},z_\zs{2})=
(y_\zs{0}(t,x,z_\zs{1},z_\zs{2}) ,c_\zs{0}(t,x,z_\zs{1},z_\zs{2}))$ with
\begin{equation}\label{4.1}
y_\zs{0}(t,x,z_\zs{1},z_\zs{2})
\,=\,\frac{z_\zs{1}}{x|z_\zs{2}|}\,\theta_\zs{t}\ \ \ \mbox{and} \ \ \
c_\zs{0}(t,x,z_\zs{1},z_\zs{2})
\,=\,\,\left(\frac{\gamma_\zs{1}}{z_\zs{1}}\right)^{q_\zs{1}}\,.
\end{equation}
Now we solve the HJB equation \eqref{A.6}, which has for our problem the following form:
\begin{equation}\label{4.2}
\left\{\begin{array}{l}
z_\zs{t}(t,x)\,+\,
r_\zs{t}\,x\,z_\zs{x}(t,x)\,+\,
\dfrac{z^2_\zs{x}(t,x)\,|\theta_\zs{t}|^2}
{2|z_\zs{xx}(t,x)|}\,
+\,\dfrac1{q_\zs{1}}\,\left(\dfrac{\gamma_\zs{1}}{z_\zs{x}(t,x)}\right)^{q_\zs{1}-1}\,=\,0\,,
\\[5mm]
z(T,x)\,=\,x^{\gamma_\zs{2}}\,.
\end{array}\right.
\end{equation}
We make the following ansatz:
\begin{equation}\label{4.3}
z(t,x)=\frac{A_\zs{1}(t)}{\gamma_\zs{1}} g^{1-q_1}(t,x)+
\frac{A_\zs{2}(t)}{\gamma_\zs{2}}g^{1-q_2}(t,x)\,,
\end{equation}
where
the function $g$ is defined in \eqref{3.3}.
One can now prove directly that this function satisfies equation \eqref{4.2} using the following properties of $g$
\beao
\Big(-A_1(t) q_1 g^{-q_1}-A_2(t) q_2 g^{-q_2}\Big)\frac{\partial}{\partial x} g(t,x)&=& g(t,x)\,,\\
\dot{A}_\zs{1}(t) g^{-q_1}(t,x)+\dot{A}_\zs{2}(t) g^{-q_2}(t,x)
-A_1(t)q_1 g^{-q_1-1}\frac{\partial}{\partial t} g(t,x)  \\
- A_2(t)q_2 g^{-q_2-1} \frac{\partial}{\partial t}g(t,x) & = & 0\\
\dot{A}_\zs{1}(t) g^{-q_1}(t,x)+\dot{A}_\zs{2}(t) g^{-q_2}(t,x)+\frac{1}{\frac{\partial}{\partial x} g(t,x)}
\frac{\partial}{\partial t} g(t,x)
&=& 0\,.
\eeao
This implies that
\begin{equation}\label{4.4}
z_\zs{t}(t,x)=-\frac{\dot{A}_\zs{1}(t)}{1-q_1}g^{1-q_1}(t,x)-
\frac{\dot{A}_\zs{2}(t)}{1-q_2}g^{1-q_2}(t,x)\,.
\end{equation}
Moreover, $z_\zs{x}(t,x)=g(t,x)$ and $z_\zs{xx}(t,x)=-g(t,x)/p(t,x)$. Equation \eqref{4.2}
implies the following differential equations for the coefficients $A_\zs{i}$:
\begin{equation}\label{4.5}
\left\{
\begin{array}{rl}
\dot{A}_\zs{1}(t)&=-\beta_\zs{1}(t)A_\zs{1}(t)-\gamma_\zs{1}^{q_1}\,,
\quad A_1(T)=0\,,\\[2mm]
\dot{A}_\zs{2}(t)&=-\beta_\zs{2}(t)A_\zs{2}(t)\,,\quad A_\zs{2}(T)=\gamma_\zs{2}^{q_2}\,.
\end{array}
\right.
\end{equation}
The solution of this system is given by the functions
\eqref{3.2} in all points of continuity of
$(\beta_\zs{i}(t))_\zs{0\le t\le T}$. We denote this set $\Gamma$. By our conditions
(all coefficients in the model \eqref{2.1} are c\`adl\`ag functions) the Lebesgue measure of $\Gamma$
is equal to $T$.
Note that conditions \eqref{2.5} and \eqref{4.5} imply that
$$
\int^T_\zs{0}\,|\dot{A}_\zs{i}(t)|\,\d t\,<\,\infty
$$
for $i=1,2$. Moreover, the definition of $g(t,x)$ in \eqref{3.3} implies that
 $g(\cdot,\cdot)$ is continuous on $[0,T]\times (0,\infty)$. Invoking
\eqref{4.4} we obtain property \eqref{A.8}. Hence condition $\H_2$  holds.\\[2mm]
Now by \eqref{4.1} we find that
$$
H(t,x,z_\zs{x}(t,x),z_\zs{xx}(t,x))\,=
\,H_\zs{0}(t,x,z_\zs{x}(t,x),z_\zs{xx}(t,x),\vartheta^*(t,x))\,,
$$
where $\vartheta^*(t,x)=(y^*(t,x),c^*(t,x))$ with
$$
y^*(t,x)= \dfrac{p(t,x)}{x}\,\theta_\zs{t}
\quad\mbox{and}\quad
c^*(t,x)=\left(\frac{\gamma_\zs{1}}{g(t,x)}\right)^{q_1}\,.
$$
Hence $\H_\zs{2}$ holds.\\[2mm]
Now we check condition $\H_\zs{3}$.
First note that equation \eqref{A.9} is identical to equation \eqref{3.5}.
By It\^o's formula one can show that this equation has a unique strong positive solution  given by
\begin{equation}\label{4.6}
X^*_\zs{t}\,=\,A_\zs{1}(t)\,g^{-q_1}(0,x)\,e^{-q_1 \xi_\zs{t}}
+A_\zs{2}(t)\,g^{-q_2}(0,x)\,e^{-q_2\xi_\zs{t}}
\end{equation}
with
$$
\xi_\zs{t}=
-\int^t_\zs{0}\Big(r_\zs{u}+\frac{1}{2}|\theta_\zs{u}|^2\Big)\d u-
\int^t_\zs{0}\theta'_\zs{u}\d W_\zs{u}\,.
$$
This implies $\H_\zs{3}$. \\[2mm]
To check the final condition $\H_\zs{4}$   note that
by definitions \eqref{3.3} and \eqref{4.6}
$$
g(t,X^*_\zs{t})=g(0,x)e^{\xi_\zs{t}}\,.
$$
Therefore, taking into account that
$$
X^*_\zs{s}\,=\,A_\zs{1}(s)\,g^{-q_1}(s,X^*_\zs{s})\,
+A_\zs{2}(s)\,g^{-q_2}(s,X^*_\zs{s})
$$
we obtain  for $s\ge t$
$$
X^*_\zs{s}\,=\,A_\zs{1}(s)\,g^{-q_1}(t,X^*_\zs{t})\,e^{-q_1(\xi_\zs{s}-\xi_\zs{t})}
+A_\zs{2}(s)\,g^{-q_2}(t,X^*_\zs{t})\,e^{-q_2(\xi_\zs{s}-\xi_\zs{t})}\,.
$$
Hence, for $s\ge t$ we can find an upper bound of the process $z(s,X^*_\zs{s})$ given by
\begin{align*}
z(s,X^*_\zs{s})&\le \frac{g(t,X^*_\zs{t})}{\min(\gamma_\zs{1},\gamma_\zs{2})}\,e^{\xi_\zs{s}-\xi_\zs{t}}\,X^*_\zs{s}
\le M_\zs{*}(X^*_\zs{t})\,
\left(e^{(1-q_1)(\xi_\zs{s}-\xi_\zs{t})}+e^{(1-q_2)(\xi_\zs{s}-\xi_\zs{t})}\right)
\,,
\end{align*}
where
$$
M_\zs{*}(x)=\frac{\sup_\zs{0\le t\le T}(A_\zs{1}(t)+A_\zs{2}(t)) \left(g^{1-q_1}(t,x)+g^{1-q_2}(t,x)\right)} {\min(\gamma_\zs{1},\gamma_\zs{2})}\\.
$$
Moreover, note that the random variables $\xi_\zs{s}-\xi_\zs{t}$ and $X^*_\zs{t}$ are independent.
Therefore, for every $m>1$  we calculate
($\E_\zs{t,x}$ is the expectation operator conditional on
$X^{\varsigma}_\zs{t}=x$)
\begin{align*}
\E_\zs{t,x}\,\sup_\zs{t\le s\le T}z^m(s,X^*_\zs{s})\le 2^{m-1}M^m_\zs{*}(x)
\left(
\E \sup_\zs{t\le s\le T}e^{m_1(\xi_\zs{s}-\xi_\zs{t})}
+
\E \sup_\zs{t\le s\le T} e^{m_2(\xi_\zs{s}-\xi_\zs{t})}
\right)\,,
\end{align*}
where $m_\zs{1}=m(1-q_1)$ and $m_\zs{2}=m(1-q_2)$.
Therefore, to check condition $\H_\zs{4}$ it suffices to show that
for every $\lambda\in\bbr$
\begin{equation}\label{4.7}
\E \sup_\zs{t\le s\le T} e^{\lambda(\xi_\zs{s}-\xi_\zs{t})}<\infty\,.
\end{equation}
Indeed, for every $t\le s\le T$ we set $
\cE_\zs{t,s}=e^{-\lambda\int^s_\zs{t}\theta'_\zs{u}\d W_\zs{u}-\frac{\lambda^2}{2}\int^s_\zs{t}|\theta_\zs{u}|^2\d u}$, then
\begin{align*}
e^{\lambda(\xi_\zs{s}-\xi_\zs{t})}\le
e^{|\lambda|R_\zs{T}+\frac{|\lambda|+\lambda^2}{2}\|\theta\|^2_\zs{T}}\cE_\zs{t,s}\,.
\end{align*}
We recall from \eqref{2.3} that
$(\theta_\zs{s})_\zs{0\le s\le T}$ is a deterministic function.
This implies that the process $(\cE_\zs{t,s})_\zs{t\le s\le T}$ is a martingale.
Hence applying the maximal inequality
for positives submartingales (see e.g. Theorem~3.2 in \cite{LiSh}) we obtain that
\begin{align*}
\E\,\sup_\zs{t\le s\le T} \cE^2_\zs{t,s}\le
4\,\E\,\cE^2_\zs{t,T}=4e^{\lambda^2\int^T_\zs{t}|\theta_\zs{u}|^2\d u}
\le 4e^{\lambda^2\|\theta\|^2_\zs{T}}\,.
\end{align*}
{}From this inequality \eqref{4.7} follows, which implies $\H_\zs{4}$.
 Therefore, by Theorem~\ref{Th.A.1} we get Theorem~\ref{Th.3.2}.
\eproof

\subsection{Proof of Theorem~\ref{Th.3.5}}
\label{subsec:4.3}

First note that restriction \eqref{2.12} is equivalent to
\begin{equation}\label{4.8}
\inf_\zs{0\le t\le T}\,L_\zs{t}(\varsigma)\,\ge\,\ln(1-\zeta)\,,
\end{equation}
where
\begin{equation}\label{4.9}
L_\zs{t}(\varsigma)\,=\,(y,\theta)_\zs{t}\,-\,V_\zs{t}\,-\, \frac{1}{2}\,\|y\|^2_\zs{t}\,
\,-\,|z_\alpha|\,\|y\|_\zs{t}
\end{equation}
with notations as in \eqref{2.3} and \eqref{2.9}.
Inequality \eqref{4.8} and the Cauchy-Schwartz inequality imply that
$$
\|y\|_\zs{T}\,\|\theta\|_\zs{T}
-\frac{1}{2}\|y\|^2_\zs{T}-|z_\alpha|\,\|y\|_\zs{T}\,\ge\,\ln(1-\zeta)
$$
and, consequently,
\begin{equation}\label{4.10}
\|y\|_\zs{T}\,\le\,\rho^*_\zs{\rm VaR}\,,
\end{equation}
where $\rho^*_\zs{\rm VaR}$ has been defined in (\ref{3.4}) and satisfies the equation
\begin{equation}\label{4.11}
\|\theta\|_\zs{T}\rho^*_\zs{\rm VaR}
-\frac{1}{2}\,(\rho^*_\zs{\rm VaR})^2-|z_\alpha|\,\rho^*_\zs{\rm VaR}\,=\,\ln(1-\zeta)\,.
\end{equation}
Moreover, for every $\varsigma\in\cU$ equation (\ref{2.8}) yields
$$
\E_\zs{x} X^{\varsigma}_\zs{t}=x\,e^{R_\zs{t}-V_\zs{t}+(y,\theta)_t}\,.
$$
For every $y\in\bbr^d$ the upper bound (\ref{4.10}) and the
Cauchy-Schwartz inequality yield
$$
\sup_\zs{0\le t\le T} e^{(y,\theta)_t} \, \le \,e^{\rho^*_\zs{\rm VaR}\|\theta\|_T}\,.
$$
Therefore,  the cost function \eqref{2.10} has an upper bound given by
\begin{align*}
J(x,\varsigma)&=x\,\left(\int^T_0\,e^{R_\zs{t}-V_\zs{t}+(y,\theta)_t}v_t \,\d t
+e^{R_\zs{T}-V_\zs{T}+(y,\theta)_T} \,\right)\\ \nonumber
&\le x\, e^{\rho^*_\zs{\rm VaR}\|\theta\|_T+R_\zs{T}}\,\left(\int^T_0 e^{-V_\zs{t}}v_t\d t+
e^{-V_\zs{T}}\right)\\ \nonumber
&= x\, e^{\rho^*_\zs{\rm VaR}\|\theta\|_\zs{T}+R_\zs{T}}\,.
\end{align*}
It is easy to see that the control $\varsigma^*$ defined in \eqref{3.12} matches
this upper bound,
i.e. $J(x,\varsigma^*)=x\,e^{\rho^*_\zs{\rm VaR}\|\theta\|_T+R_\zs{T}}$.
To finish the proof we have to check condition (\ref{4.8}) for this control.
If $\|\theta\|_\zs{T}=0$ then by \eqref{4.9}
\begin{align*}
L_\zs{t}(\varsigma^*)&=-\frac{1}{2}\,\|y^*\|^2_\zs{t}-|z_\alpha|\,\|y^*\|_\zs{t}
\ge
-\frac{1}{2}\,\|y^*\|^2_\zs{T}-|z_\alpha|\,\|y^*\|_\zs{T}\\
&\ge
-\frac{1}{2}(\rho^*_\zs{\rm VaR})^2
-|z_\alpha|\,\rho^*_\zs{\rm VaR}=\ln(1-\zeta)\,.
\end{align*}
Let now $\|\theta\|_\zs{T}>0$. Note that condition \eqref{3.10} implies
$|z_\zs{\alpha}|\ge 2\|\theta\|_\zs{T}-\rho^*_\zs{\rm VaR}$. Moreover,
we can represent $L_\zs{t}(\varsigma^*)$ as
$$
L_\zs{t}(\varsigma^*)\,=\,\rho^*_\zs{\rm VaR}\,f\left(\|\theta\|_\zs{t}/\|\theta\|_\zs{T}\right)
$$
with
$$
f(\eta)=
(2\,\|\theta\|_\zs{T}-\,\rho^*_\zs{\rm VaR}\,)\,\frac{\eta^2}{2}\,
-|z_\alpha|\,\eta\,,\quad 0\le\eta\le 1\,.
$$
Then
$$
\inf_\zs{0\le t\le T}
L_\zs{t}(\varsigma^*)\,=\,\rho^*_\zs{\rm VaR}\,\inf_\zs{0\le \eta\le 1}\,f(\eta)\,.
$$
Taking into account that for $|z_\zs{\alpha}|\ge 2\|\theta\|_\zs{T}-\rho^*_\zs{\rm VaR}$
this infimum equals $f(1)$ we obtain together with (\ref{4.11})
$$
\inf_\zs{0\le t\le T}
L_\zs{t}(\varsigma^*)=\rho^*_\zs{\rm VaR}f(1)=\ln(1-\zeta)\,.
$$
This proves Theorem~\ref{Th.3.5}.
\eproof

\subsection{Proof of Theorem~\ref{Th.3.9}}
\label{subsec:4.5}

We have to prove condition \eqref{4.8} for the strategy \eqref{3.7}--\eqref{3.8}:
\begin{align*}
L_\zs{t}(\varsigma^*)&\,=\,\left(q\,-\,\frac{q^2}{2}\right)\,\|\theta\|^2_\zs{t}\,-\,V^*_\zs{t}
\,-\,q\,|z_\alpha|\,\|\theta\|_\zs{t}\\
&\ge\,\left(q\,-\,\frac{q^2}{2}\right)\,\|\theta\|^2_\zs{T}\Chi_\zs{\{q>2\}}
-\,V^*_\zs{T}
\,-\,q\,|z_\alpha|\,\|\theta\|_\zs{T}=l_*(\gamma)\,.
\end{align*}
Now condition \eqref{4.8} follows immediately from the restrictions on $\zeta$
and the definition of $l_*(\gamma)$.
\eproof

\subsection{Proof of Theorem~\ref{Th.3.12}}\label{subsec:4.6}

We prove this theorem as theorem~\ref{Th.3.5}. Firstly, we find
an upper bound for the cost function $J(x,\varsigma)$ and, secondly, we show that
the optimal control \eqref{3.26} matches this bound and satisfies
condition \eqref{4.8}.
To this end  note that from \eqref{2.8} we find that for $\varsigma\in \cU$
\begin{equation}\label{4.12}
\E_x\,(X^\varsigma_\zs{t})^{\gamma}=x^{\gamma}\,\wh{g}_\zs{\gamma}(t)
e^{-\gamma V_\zs{t}+\gamma (y,\theta)_\zs{t}-
\frac{\gamma(1-\gamma)}{2}\|y\|^{2}_\zs{t}}\,.
\end{equation}
This implies for $\varsigma\in\cU$ that
 the cost function \eqref{2.10} has the form
$$
J(x,\varsigma)\,=\,x^{\gamma_\zs{1}}\,\int^T_0(e^{-V_\zs{t}}v_t)^{\gamma_\zs{1}}\,\wh{g}_\zs{1}(t)
\,\wh{h}_\zs{1}(t,y)\d t+x^{\gamma_\zs{2}}\,\wh{g}_\zs{2}(T)\,
e^{-\gamma_\zs{2} V_\zs{T}}\,\wh{h}_\zs{2}(T,y)\,,
$$
where
$$
\wh{h}_\zs{i}(t,y)=
e^{\gamma_\zs{i} (y,\theta)_\zs{t}-\frac{\gamma_\zs{i}(1-\gamma_\zs{i})}{2}\|y\|^2_\zs{t}}\,.
$$
H\"older's inequality with $p=1/\gamma_\zs{1}$ and $q=(1-\gamma_\zs{1})^{-1}$ yields
\begin{align*}
J(x,\varsigma)
&\le\,\sup_\zs{0\le t\le T}\,\wh{h}(t,y)
\left(x^{\gamma_\zs{1}}\int^T_0(e^{-V_\zs{t}}v_t)^{\gamma_\zs{1}} \wh{g}_\zs{1}(t)
\d t+x^{\gamma_\zs{2}}\wh{g}_\zs{2}(T)
e^{-\gamma_\zs{2} V_\zs{T}} \right) \\
&\le\,\sup_\zs{0\le t\le T}\wh{h}(t,y)
\left(x^{\gamma_\zs{1}} (1-e^{-V_\zs{T}})^{\gamma_\zs{1}}\,\|\wh{g}_\zs{1}\|_{q,T}
+ x^{\gamma_\zs{2}}\wh{g}_\zs{2}(T)\,e^{-\gamma_\zs{2} V_\zs{T}}\right)\,,
\end{align*}
where
$\wh{h}(t,y)=\max\{\wh{h}_\zs{1}(t,y),\wh{h}_\zs{2}(t,y)\}$.
We abbreviate as before
$\|\wh{g}_\zs{1}\|_{q,T}:=(\int_0^T e^{q \gamma_\zs{1} R_t} \d t)^{1/q}$.
By setting  $\kappa=1-e^{-V_\zs{T}}$ we obtain that
\begin{equation}\label{4.13}
J(x,\varsigma)
\le\,\max_\zs{0\le t\le T}\,\wh{h}(t,y)\,G(x,\kappa)\,,
\end{equation}
where  $G(\cdot,\cdot)$ is given in \eqref{3.22}.
Moreover, condition \eqref{4.8} implies
\begin{equation}\label{4.14}
\|y\|_\zs{T} \le
\sqrt{\left(|z_\zs{\alpha}|-\|\theta\|_\zs{T}\right)^2+2\ln\frac{1-\kappa}{1-\zeta}}-
(|z_\zs{\alpha}|-\|\theta\|_\zs{T}):=\varrho(\kappa)
\end{equation}
and $0\le \kappa \le \zeta <1$.
It is easy to see that $\varrho(\kappa)\le \varrho(0)=\rho^*_\zs{\rm VaR}$ for every
$0\le \kappa\le \zeta$.
From this inequality follows that for $i=1,2$
the functions $\wh{h}_\zs{i}(t,y)$ with $0<\gamma_\zs{i}\le 1$ can be bounded above by
\begin{align}\nonumber
\sup_\zs{0\le t\le T}\,
\wh{h}_\zs{i}(t,y)\,&\le
\exp\left\{\gamma_\zs{i}\max_{0\le x\le \varrho(\kappa)}\left( x\|\theta\|_\zs{T}-
\frac{(1-\gamma_\zs{i})x^2}{2}\right)\right\}\\ \label{4.15}
&=\exp\{\gamma_\zs{i} \varrho_\zs{i}(\kappa)\|\theta\|_\zs{T}-
\frac{\gamma_\zs{i}(1-\gamma_\zs{i})}{2}\varrho^2_\zs{i}(\kappa)\}
:=M_\zs{i}(\varrho_\zs{i}(\kappa))\,,
\end{align}
where $\varrho_\zs{i}(\kappa)=\min(\varrho(\kappa),x_\zs{i})$ with
$x_\zs{i}=q_\zs{i}\,\|\theta\|_\zs{T}$ for $0<\gamma_\zs{i}<1$ and
$\varrho_\zs{i}(\kappa)=\varrho(\kappa)$ for  $\gamma_\zs{i}=1$.
Therefore, from (\ref{4.13}) we obtain the following upper
 bound for the cost function
\begin{align*}
J(x,\varsigma)&
\le\,\max_\zs{1\le i\le 2}\,M_\zs{i}(\varrho_\zs{i}(\kappa))G(x,\kappa)\,
\le\,\max_\zs{1\le i\le 2}\,
\sup_\zs{0\le \kappa\le \zeta}\,M_\zs{i}(\varrho_\zs{i}(\kappa))\,G(x,\kappa)\,.
\end{align*}
If $\varrho(0)\le x_\zs{i}$ then
$$
\sup_\zs{0\le\kappa\le\zeta}\,M_\zs{i}(\varrho_\zs{i}(\kappa))\,G(x,\kappa)=
\sup_\zs{0\le\kappa\le\zeta}\,M_\zs{i}(\varrho(\kappa))\,G(x,\kappa)\,.
$$
We calculate this supremum by means of Lemma~\ref{Le.A.1} with $a=0$ and $b=\zeta$.
Note that condition \eqref{3.24} guarantees that $\zeta<\kappa_\zs{*}(x)$, which is defined in \eqref{3.23}.
Therefore, the function $G(x,\cdot)$ has positive first derivative and
 negative second on $[0,\zeta]$. Moreover, from \eqref{4.14} we find
 the derivative of $\varrho(\cdot)$ as
$$
\dot{\varrho}(\kappa)\,=\,-\,
\frac{1}{(1-\kappa)\sqrt{(|z_\zs{\alpha}|-\|\theta\|_\zs{T})^2+2\ln(1-\kappa)-2\ln(1-\zeta)}}
$$
and, therefore,
$$
\sup_\zs{0\le\kappa\le\zeta}\,
|\dot{\varrho}(\kappa)|\,
\le \frac{1}{(1-\zeta)(|z_\zs{\alpha}|-\|\theta\|_\zs{T})}\,.
$$
By \eqref{3.25} we obtain that
$$
\sup_\zs{0\le \kappa\le \zeta}\,|\dot{\rho}(\kappa)|\le\,
\frac{1}{\max\{\gamma_\zs{1}\,,\,\gamma_\zs{2}\}\,\|\theta\|_\zs{T}}\,
\frac{\partial\ln G(x,\zeta)}{\partial\zeta}\,.
$$
Now Lemma~\ref{Le.A.1} yields
\begin{equation}\label{4.16}
\max_\zs{0\le\kappa\le \zeta}M_\zs{i}(\varrho(\kappa))\,G(x,\kappa)\,
=\,M_\zs{i}(\varrho(\zeta))G(x,\zeta)=G(x,\zeta)\,.
\end{equation}
Consider now $x_\zs{i}<\varrho(0)$. We recall that
$\varrho(\cdot)$ is decreasing on $[0,\zeta]$ with $\varrho(\zeta)=0$.
Therefore,  there exists $0\le\kappa_\zs{i}<\zeta$ such that $\varrho(\kappa_\zs{i})=x_\zs{i}$.
As $G(x,\cdot)$ is increasing on $[0,\zeta]$ we obtain
\begin{align*}
\max_\zs{0\le\kappa\le\kappa_\zs{i}}\,M_\zs{i}(\varrho_\zs{i}(\kappa))\,G(x,\kappa)\,=\,
M_\zs{i}(\varrho(\kappa_\zs{i}))\,G(x,\kappa_\zs{i})\,.
\end{align*}
This in combination with \eqref{4.16} yields
$$
\sup_\zs{0\le\kappa\le\zeta}M_\zs{i}(\varrho_i(\kappa))\,G(x,\kappa)\,=
\sup_\zs{\kappa_\zs{i}\le\kappa\le\zeta}M_\zs{i}(\varrho(\kappa))\,G(x,\kappa)\,
=\,G(x,\zeta)\,.
$$
This implies the following upper bound for the cost function
\begin{equation}\label{4.17}
J(x,\varsigma)\le G(x,\zeta)\,.
\end{equation}
Now we find a control to obtain the equality in \eqref{4.17}.
It is clear that we have to take a consumption such that
$$
\int^T_0\,\wh g_\zs{1}(t)(e^{-V_\zs{t}}v_t)^{\gamma_\zs{1}} \d t=
(1-e^{-V_\zs{T}})^{\gamma_\zs{1}}\,\|\wh g_\zs{1}\|_{q_\zs{1},T}
$$
and
$V_\zs{T}=-\ln(1-\zeta)$. To find this consumption
we solve the differential equation on $[0,T]$
$$
\dot{V}_\zs{t}\,e^{-V_t}\,=\,
\frac{\zeta}{\|\wh g_\zs{1}\|^{q_\zs{1}}_\zs{q_\zs{1},T}}\,
 \wh g^{q_\zs{1}}_\zs{1}(t) \,, \quad V_\zs{0}=0\,.
$$
The solution of this equation is given by
$$
V^*_\zs{t}=-
\ln
\left(
1-
\zeta
\frac{\|\wh{g}_\zs{1}\|^{q_\zs{1}}_\zs{q_\zs{1},t}}{\|\wh{g}_\zs{1}\|^{q_\zs{1}}_\zs{q_\zs{1},T}}
\right)
$$
and the optimal consumption rate is
$$
v^*_\zs{t}=\dot{V}^*_\zs{t}=
\frac{\zeta\,\wh{g}^{q_\zs{1}}_\zs{1}(t)}
{\|\wh{g}_\zs{1}\|^{q_\zs{1}}_\zs{q_\zs{1},T}-\zeta\|\wh{g}_\zs{1}\|^{q_\zs{1}}_\zs{q_\zs{1},t}}\,.
$$
We recall that $r_\zs{t}\ge 0$, therefore, for every $0\le t\le T$
$$
v^*_\zs{t}\le v^*_\zs{T}=
\frac{\zeta\wh{g}^{q_\zs{1}}_\zs{1}(T)}{(1-\zeta)\|\wh{g}_\zs{1}\|^{q_\zs{1}}_\zs{q_\zs{1},T}}\,.
$$
The condition $0< \zeta\le \wh\kappa(\gamma_\zs{1})$ implies directly that
the last upper bound less than $1$, i.e. the strategy $\varsigma^*$ defined in
\eqref{3.27}
belongs to $\cU$.
Moreover, from \eqref{4.14} we see that for the value
 $V^*_\zs{T}=-\ln(1-\zeta)$ (i.e. $\kappa=\zeta$) the only control process, which satisfies this condition is identical zero; i.e.
$y^*_\zs{t}=0$ for all $0\le t\le T$. In this case $\wh{h}(t,y^*)=1$ for every
  $t\in [0,T]$ and, therefore, $J(x,\varsigma^*)=G(x,\zeta)$.
\eproof

\subsection{Proof of Lemma~\ref{Le.3.15}}
\label{subsec:4.7}

(1) \, Recall the following well known inequality for the Gaussian integral
\begin{equation}\label{4.18}
(1-x^{-2})e^{-x^2/2}<
x\int^{\infty}_\zs{x}\,e^{-t^2/2}\,\d t< e^{-x^2/2}\,,\quad x\ge 0\,.
\end{equation}
We use this to check directly that $\psi(\rho,\cdot)$ is for every fixed $\rho>0$
decreasing for $|z_\zs{\alpha}|\,\ge 2\|\theta\|_\zs{T}$.
This implies for $0\le u\le 1$
\begin{align*}
\frac{\partial \psi(\rho,u)}{\partial u} =
2\|\theta\|_\zs{T}\,\rho\,u\,-\,\rho\,
\frac{e^{-(|z_\zs{\alpha}|\,+\rho\,u)^2/2}}
{\int^{\infty}_\zs{|z_\zs{\alpha}|\,+\,\rho\,u }\,e^{-t^2/2}\,\d t}
\le\,\rho\,(2\|\theta\|_\zs{T}\,-\,|z_\zs{\alpha}|)\,<\,0\,.\nonumber
\end{align*}
(2) \, Similarly, we can show that $\psi(\cdot,1)$ is strictly decreasing
for $|z_{\alpha}|\ge \|\theta\|_\zs{T}$.\\
(3) \, From (\ref{4.18}) we obtain
\begin{align}
\psi(\rho,1) 
&\le\,
\|\theta\|_\zs{T}\,\rho\,-\,
\ln\,\int^{\infty}_{|z_\zs{\alpha}|}\,e^{-t^2/2}\,\d t
-\frac{1}{2}\,(|z_\zs{\alpha}|+\rho)^2-\ln(|z_\zs{\alpha}|+\rho)\,\label{4.19}
\end{align}
This implies that $\lim_\zs{\rho\to\infty}\psi(\rho,1)=-\infty$.
As $\psi(0,1)=0$ we conclude that
the equation $\psi(\rho,1)=a$
has a unique root for every $a\le 0$.
Thus $\rho^*_{\rm ES}$ is
equal to the root of this equation for  $a=\ln(1-\zeta)$.
Now for $|z_\zs{\alpha}|>1$
inequalities \eqref{4.18}--\eqref{4.19}
imply directly the upper bound for $\rho^*_{\rm ES}$
as given in \eqref{rho3}.
\eproof

\subsection{Proof of Theorem~\ref{Th.3.16}}
\label{subsec:4.8}

Note that Lemma~\ref{Le.3.15} implies immediately that $\rho^*_{\rm ES}<\infty$
and $\psi(\rho^*_{\rm ES},1)=\ln(1-\zeta)$.
Furthermore, inequality \eqref{2.13} is equivalent to
\begin{equation}\label{4.20}
\inf_\zs{0\le t\le T}\,
L^*_\zs{t}(\varsigma)\,\ge\,\ln(1-\zeta)\,,
\end{equation}
where
$$
L^*_\zs{t}(\varsigma)=(y\,,\theta)_\zs{t}-V_\zs{t}\,
+\ln\,(F_\zs{\alpha}(|z_\zs{\alpha}|+\|y\|_\zs{t}))\,.
$$
First note that
\begin{align*}
L^*_\zs{T}(\varsigma)&=(y\,,\theta)_\zs{T}-V_\zs{T}\,
+\ln\,(F_\zs{\alpha}(|z_\zs{\alpha}|+\|y\|_\zs{T}))\\
&\le \|y\|_\zs{T}\,\|\theta\|_\zs{T}
+\ln\,(F_\zs{\alpha}(|z_\zs{\alpha}|+\|y\|_\zs{T}))
=\psi(\|y\|_\zs{T},1)
\,.
\end{align*}
Therefore, for every strategy $\varsigma\in\cU$
satisfying inequality \eqref{4.20}
for $t=T$ we obtain
\begin{align*}
\ln(1-\zeta)=\psi(\rho^*_{\rm ES},1)\le L^*_\zs{T}(\varsigma)\le \psi(\|y\|_\zs{T},1)\,.
\end{align*}
By Lemma~\ref{Le.3.15}(2)  $\psi(\cdot,1)$ is decreasing, hence
$\|y\|_\zs{T}\le  \rho^*_{\rm ES}$. Therefore,
to conclude the proof we have to show (\ref{4.20})
for the strategy $\varsigma^*$ as defined in
\eqref{3.12} with $\rho^*_\zs{\rm VaR}=\rho^*_{\rm ES}$. \\
If $\|\theta\|_\zs{T}=0$, then $\varsigma^*=(y^*,0)$ with every function $y^*$ for which
$\|y^*\|_\zs{T}\le \rho^*_{\rm ES}$.
Therefore, if $\|\theta\|_\zs{T}=0$, then
$$
L^*_\zs{t}(\varsigma^*)=
\psi(\|y^*\|_\zs{t},1)\ge
\psi(\|y^*\|_\zs{T},1)\,\ge\,\ln(1-\zeta)\,.
$$
If $\|\theta\|_\zs{T}>0$, then
\begin{align*}
\inf_\zs{0\le t\le T}
L^*_\zs{t}(\varsigma^*)\,&=\inf_\zs{0\le t\le T}
\psi\left(\rho^*_{\rm ES},\frac{\|\theta\|_\zs{t}}{\|\theta\|_\zs{T}}\right)
=
\psi(\rho^*_{\rm ES},1)\,=\,\ln(1-\zeta)\,.
\end{align*}
This proves Theorem~\ref{Th.3.16}.
\eproof

\subsection{Proof of Theorem~\ref{Th.3.18}}
\label{subsec:4.10}

It suffices to prove condition \eqref{4.20}
for the strategy \eqref{3.7}--\eqref{3.8}. We have
\begin{align}\nonumber
L^*_\zs{t}(\varsigma^*)&\,=\,\int^t_\zs{0}\,(y^*_\zs{u})'\,\theta_\zs{u}\,\d u\,-
\,V^*_\zs{t}\,+\,
\ln\,(F_\zs{\alpha}(|z_\zs{\alpha}|+\|y^*\|_\zs{t}))\\[2mm] \nonumber
&=q\|\theta\|^2_\zs{t}\,-
\,V^*_\zs{t}\,+\,
\ln\,(F_\zs{\alpha}(|z_\zs{\alpha}|\,+\,q\,\|\theta\|_\zs{t}))\\[2mm] \label{4.21}
&\ge \psi_\zs{0}(\|\theta\|_\zs{t})-V^*_\zs{T}\,,
\end{align}
where
$$
\psi_\zs{0}(u)\,=q\,u^2\,+\,
\ln\,F_\zs{\alpha}\,(|z_\zs{\alpha}|\,+\,q\,u)
\quad\mbox{with}\quad q=\frac{1}{1-\gamma}\,.
$$
It is clear that $\psi_\zs{0}$ is continuously differentiable.
 Moreover,  by inequality \eqref{4.18} we obtain
 for $0\le u\le \|\theta\|_\zs{T}$
\begin{align*}
\frac{\d \psi_\zs{0}(u)}{\d u}&\,=\,2\,q\,u\,-\,q\,
\frac{e^{-(|z_\zs{\alpha}|\,+\,q\,u)^2/2}}
{\int^{\infty}_\zs{|z_\zs{\alpha}|\,+q u}\,e^{-t^2/2}\,\d t}\\[2mm]
&
\le\,2\,\,q\,u\,-\,q\,
|z_\zs{\alpha}|\,-\,q^2\,u\,
\le\,q\,(2\|\theta\|_\zs{T}\,-\,|z_\zs{\alpha}|)\,.
\end{align*}
Since $|z_\zs{\alpha}|\ge 2\|\theta\|_\zs{T}$, $\psi_\zs{0}(u)$
decreases in $[0,\|\theta\|_\zs{T}]$.
Hence, inequality \eqref{4.21} implies
\begin{align*}
L^*_\zs{t}(\varsigma^*)\ge\,\psi_\zs{0}(\|\theta\|_\zs{T})-V^*_\zs{T}=q\|\theta\|^2_\zs{T}\,+\,
\ln e^{-V^*_\zs{T}}\,F_\zs{\alpha}(|z_\zs{\alpha}|+q\,\|\theta\|_\zs{T})\,.
\end{align*}
Applying  condition \eqref{3.32} yields \eqref{4.20}.
This proves Theorem~\ref{Th.3.18}.
\eproof

\subsection{Proof of Theorem~\ref{Th.3.19}}
\label{subsec:4.11}

We recall that $\psi(\rho,1)\le 0$ for $\rho\ge 0$. Therefore
condition \eqref{4.20} implies
\begin{equation}\label{4.22}
\ln(1-\zeta)\le-V_\zs{T}+\psi(\|y\|_\zs{T},1)\le -V_\zs{T}\,.
\end{equation}
As in the proof of
Theorem~\ref{Th.3.12} we set
$\kappa=1-e^{-V_\zs{T}}$ and conclude from this inequality that $0\le\kappa\le\zeta$.
Moreover, from \eqref{4.22} we obtain also that
$$
\ln(1-\zeta)-\ln(1-\kappa)\le\psi(\|y\|_\zs{T},1)\,.
$$
Since, by Lemma~\ref{Le.3.15}(2) $\psi(\cdot,1)$ is decreasing, we get
$\|y\|_\zs{T}\le \rho(\kappa)$,
where $\rho(\kappa)$ is the solution of the equation
\begin{equation}\label{4.23}
\psi(\rho,1)=\ln(1-\zeta)-\ln(1-\kappa)\,.
\end{equation}
By Lemma~\ref{Le.3.15}(3) the root of \eqref{4.23} exists for every
$0 \le \kappa\le \zeta$ and is decreasing
in $\kappa$ giving $\rho(\kappa)\le \rho(0)=\rho^*_{\rm ES}$.
Consequently, we estimate the cost function as
in Section~\ref{subsec:4.6}
and obtain
\begin{equation}\label{4.24}
J(x,\varsigma)\le\,\max_\zs{1\le i\le 2}  \,
\max_\zs{\kappa\in [0\,,\,\zeta]}\,M_\zs{i}(\rho_\zs{i}(\kappa))\,G(x,\kappa)\,,
\end{equation}
where $G(x,\kappa)$ is as in \eqref{3.22}, $M_\zs{i}(\cdot)$ is defined in \eqref{4.15} and
$\rho_\zs{i}(\kappa)=\min(x_\zs{i},\rho(\kappa))$ for
$x_\zs{i}=\|\theta\|_\zs{T}/(1-\gamma_\zs{i})$ for $0< \gamma_\zs{i}<1$ with
$\rho_\zs{i}(\kappa)=\rho(\kappa)$ for $\gamma_\zs{i}=1$.
\\[2mm]
To finish the proof we have to show condition \eqref{A.0} of Lemma~\ref{Le.A.1}. {}From \eqref{4.23} we find that
$$
\dot{\rho}(\kappa)=\frac{1}{1-\kappa}\,
\left(\frac{\d \psi(\rho,1)}{\d \rho}\right)^{-1}\,.
$$
Now from the definition of $\psi$ in \eqref{3.29}
and  inequality \eqref{4.18} follows
$$
\frac{\d \psi(\rho,1)}{\d \rho} =
\|\theta\|_\zs{T}\,-\,
\frac{e^{-(|z_\zs{\alpha}|\,+\,\rho)^2/2}}
{\int^{\infty}_\zs{|z_\zs{\alpha}|\,+\,\rho}\,e^{-t^2/2}\,\d t}\le
 \,\|\theta\|_\zs{T}-\,|z_\zs{\alpha}|\,.
$$
Therefore \eqref{3.33} yields (we set $G_1(x,\zeta)=\frac{\partial G_(x,\zeta)}{\partial\zeta}$)
$$
\sup_\zs{0\le \kappa\le \zeta}
|\dot{\rho}(\kappa)|\le \frac{1}{(1-\zeta)(|z_\zs{\alpha}|-\|\theta\|_\zs{T})}
\,\le\,
\frac{G_\zs{1}(x,\zeta)}{\max\{ \gamma_\zs{1}\,,\,\gamma_\zs{2}\}\|\theta\|_\zs{T}G(x,\zeta)}\,.
$$
We apply Lemma~\ref{Le.A.1}, and the same reasoning as in the proof of Theorem~\ref{Th.3.12}
implies that
$$
\max_\zs{0\le\kappa\le \zeta}\,M_\zs{i}(\rho_\zs{i}(\kappa))\,G(x,\kappa)\le
G(x,\zeta)
$$
for $i=1,2$. Therefore from the upper bound \eqref{4.24} follows
$$
J(x,\varsigma)\le\,G(x,\zeta)\,.
$$
The remainder of the proof is the same as for Theorem~\ref{Th.3.12}.
\eproof

\renewcommand{\theequation}{A.\arabic{equation}}
\renewcommand{\thetheorem}{A.\arabic{theorem}}
\renewcommand{\thesubsection}{A.\arabic{subsection}}
\section*{Appendix}
\setcounter{equation}{0}
\setcounter{theorem}{0}
\setcounter{subsection}{0}
\addcontentsline{toc}{section}{Appendix}

\subsection{A Technical Lemma}
\label{subsec:A.1}

\begin{lemma}\label{Le.A.1}
Let $G$ be some positive two times continuously differentiable function
on $[a,b]$ such that $\dot{G}(x)\ge 0$ and $\ddot{G}(x)\le 0$ for all $a\le x\le b$. Moreover,
let \mbox{$\varrho : [a,b]\to \bbr_\zs{+}$} be continuously differentiable
with negative derivative $\dot\rho$ satisfying
\begin{equation}\label{A.0}
\sup_\zs{a\le \kappa\le b}\,|\dot{\rho}(\kappa)|\,\le
\frac{\left(\ln G(b)\right)'}{\max\{\gamma_\zs{1}\,,\,\gamma_\zs{2}\}\|\theta\|_\zs{T}}\,.
\end{equation}
Recall the definitions of $M_\zs{i}(\cdot)$ in \eqref{4.15}.
Then the functions $M_\zs{1}(\rho(\cdot))\,G(\cdot)$ and $M_\zs{2}(\rho(\cdot))\,G(\cdot)$ are increasing in $[a,b]$.
\end{lemma}

\proof
For $\|\theta\|_\zs{T}=0$ the result is obvious.
Consider  now $\|\theta\|_\zs{T}>0$.
We prove that for $i=1,2$ the functions
$l_\zs{i}(x)=\ln M_\zs{i}(\rho(x))+ \ln G(x)$ are increasing in $[a,b]$.
As derivative we obtain
$$
\dot{l}_\zs{i}(\kappa)=\gamma_\zs{i} \dot{\varrho}(\kappa)
\left(\|\theta\|_\zs{T}-(1-\gamma_\zs{i})\varrho(\kappa)\right)
+\frac{\dot{G}(x)}{G(x)}\,.
$$
Since the derivative  of the function $\dot{G}(\cdot)/G(\cdot)$
is negative on $[a,b]$,
$\dot{G}(\cdot)/G(\cdot)$  is  decreasing on $[a,b]$,
hence
$$
\frac{\dot{G}(x)}{G(x)}
\ge
\frac{\dot{G}(b)}{G(b)}\,>\,0
$$
for $x\in[a,b]$.
Therefore, as $\varrho>0$ and $\dot{\varrho}<0$ we find
$$
\dot{l}_\zs{i}(x)
\ge
\left(\ln G(b)\right)'
-
\gamma_\zs{i}\, \|\theta\|_\zs{T}\,|\dot{\varrho}(\kappa)| \ge 0\,,\quad a\le\kappa\le b\,.
\eqno{\Box}
$$

\subsection{The Verification Theorem}
\label{subsec:A.3}

We prove a special form of the verification theorem
(see e.g. Touzi~\cite{To}, p.~16).
Consider on the inteval $[0,T]$ the stochastic control process given by the It\^o process
\begin{equation}\label{A.1}
\d X^{\varsigma}_\zs{t}\,=\,a(t,X^{\varsigma}_\zs{t},\varsigma_\zs{t})\,\d t\,+\,
b(t,X^{\varsigma}_\zs{t},\varsigma_\zs{t})\,\d W_\zs{t}\,,\quad t\ge0\,,\quad
X^{\varsigma}_\zs{0}=x>0\,.
\end{equation}
We assume that the control process
$\varsigma$ takes values in some set $\cK\subseteq\bbr^d\times [0,\infty)$.
Moreover, assume that the coefficients  $a$ and $b$ satisfy the following conditions
\begin{itemize}
\item[(1)] \
for all $t\in[0,T]$ the functions
$a(t,\cdot,\cdot)$ and $b(t,\cdot,\cdot)$
are continuous on $(0,\infty)\times\cK$;
\item [(2)] \
for every deterministic vector $\upsilon\in\cK$
the stochastic differential equation
$$
\d X^{\upsilon}_\zs{t}\,=\,a(t,X^{\upsilon}_\zs{t},\upsilon)\,\d t\,+\,
b(t,X^{\upsilon}_\zs{t},\upsilon)\,\d W_\zs{t}
\,,\quad X^{\upsilon}_\zs{0}=x> 0,
$$
 has an unique strong solution.
\end{itemize}
Now we introduce  admissibles control processes for the equation
\eqref{A.1}.
 We set
$\cF_\zs{t}=\sigma\{W_\zs{u}\,,0\le u\le t\}$ for any $0<t\le T$.
\begin{definition}\label{De.A.1}
A stochastic control process
$\varsigma=(\varsigma_\zs{t})_\zs{0\le t\le T}=((y_\zs{t},c_\zs{t}))_\zs{0\le t\le T}$ is called
{\em admissible}
on $[0,T]$ with respect to equation \eqref{A.1}
if it is $(\cF_\zs{t})_\zs{0\le t\le T}$ -
progressively measurable with values in $\bbr^d\times [0,\infty)$,
and equation \eqref{A.1} has a unique strong a.s. positive continuous solution
$(X^{\varsigma}_\zs{t})_\zs{0\le t\le T}$ on $[0\,,\,T]$
such that
\begin{equation}\label{A.2}
\int^T_0\,\left(|a(t,X^{\varsigma}_\zs{t},\varsigma_\zs{t})|\,+\,
b^2(t,X^{\varsigma}_\zs{t},\varsigma_\zs{t})\right)\d t\,<\,\infty\quad \mbox{a.s..}
\end{equation}
\end{definition}
In this context $\cV$ is the set of all admissible control processes
with respect to the equation \eqref{A.1}; cf. Definition~\ref{De.2.1}.

Moreover, assume that
$f\,:\,[0,T]\,\times\,(0,\infty)\,\times\,\cK\,\to\,[0,\infty)$
and  $h\,:\,(0,\infty)\,\to\,[0,\infty)$ are continuous utility functions.
We define the cost function by
$$
J(t,x,\varsigma)\,:=\,
\E_\zs{t,x}\,\left[\int^T_\zs{t}\,f(s,X^{\varsigma}_\zs{s},\varsigma_\zs{s})\,\d s\,
+\,h(X^{\varsigma}_\zs{T})\right]\,,\quad 0\le t\le T\,,
$$
where $\E_\zs{t,x}$ is the expectation operator conditional on
$X^{\varsigma}_\zs{t}=x$.
Our goal is to solve the optimization problem
\begin{equation}\label{A.4}
J^*(t,x)\,:=\,\sup_\zs{\varsigma\in\cV}\,J(t,x,\varsigma)\,.
\end{equation}
To this end we introduce the Hamilton function
\begin{equation}\label{A.5}
H(t,x,z_\zs{1},z_\zs{2})\,:=\,\sup_\zs{\vartheta\in\cK}\,
H_\zs{0}(t,x,z_\zs{1},z_\zs{2},\vartheta)\,,
\end{equation}
where
$$
H_\zs{0}(t,x,z_\zs{1},z_\zs{2},\vartheta)\,:=\,
a(t,x,\vartheta)\,z_\zs{1}\,+\,
\frac{1}{2}\,b^2(t,x,\vartheta)\,z_\zs{2}\,+\,
f(t,x,\vartheta)\,.
$$
In order to find the solution to \eqref{A.4} we investigate the
Hamilton-Jacobi-Bellman equation
\begin{equation}\label{A.6}
\left\{\begin{array}{ll}
z_\zs{t}(t,x)\,+\,H(t,x,z_\zs{x}(t,x),z_\zs{xx}(t,x))\,=\,0\,,\quad
& \ t\in [0,T]\,,
\\[5mm]
z(T,x)\,=\,h(x)\,,\ &    x>0\,.
\end{array}\right.
\end{equation}
Here $z_t$ denotes the  partial derivative of $z$ with respect to $t$, analogous notation applies to all partial derivatives.

We assume that the following conditions hold:\\[3mm]
$\H_\zs{1})$ {\em There exists some function
$z\,:\,[0,T]\times(0,\infty)\to [0,\infty)$,
which satisfies the following conditions.
\begin{itemize}
\item
For all $0\le t_1,t_2\le T$ there exists a $\cB[0,T]\otimes\cB(0,\infty)$ measurable
function  $z_\zs{t}(\cdot,\cdot)$
such that
\begin{equation}\label{A.7}
z(t_2,x)-z(t_1,x)=\int^{t_2}_\zs{t_2}\,z_\zs{t}(u,x)\,\d u\,,\quad x>0\,.
\end{equation}
\item
Moreover, we assume that for every $u\in [0,T]$ the function
$z_\zs{t}(u,\cdot)$ is continuous on $(0,\infty)$ such that
for every $N>1$
\begin{equation}\label{A.8}
\lim_\zs{\epsilon\to 0}\,
\int^T_\zs{0}\,\sup_\zs{x,y\in K_\zs{N}\,,\,|x-y|<\epsilon}|z_\zs{t}(u,x)-z_\zs{t}(u,y) |\,\d u=0\,,
\end{equation}
where $K_\zs{N}=[N^{-1},N]$.
\item    The function $z$ has second partial derivative $z_\zs{xx}$,
which is continuous on  $[0,T]\times (0,\infty)$.
\item   There exists a set $\Gamma\subset[0,T]$ of Lebesgue measure $\la(\Gamma)=T$
such that $z(t,x)$ satisfies equation \eqref{A.6} for all $t\in\Gamma\subset [0,T]$
and for all $x>0$.
\end{itemize}
}
\noindent $\H_\zs{2})$ {\em There exists a measurable function
$\vartheta^*\,:\,[0,T]\times (0,\infty) \to \cK$
 such that
$$
H(t,x,z_\zs{x}(t,x),z_\zs{xx}(t,x))\,=\,
H_\zs{0}(t,x,z_\zs{x}(t,x),z_\zs{xx}(t,x),\vartheta^*(t,x))
$$
for all $t\in \Gamma$ and for all $x\in (0,\infty)$.}\\[3mm]
$\H_\zs{3})$ {\em There exists a unique a.s. strictly positive strong solution
to the It\^o equation
\begin{equation}\label{A.9}
\d X^*_\zs{t}\,=\,a^*(t,X^*_\zs{t})\,\d t\,+\,
b^*(t,X^*_\zs{t})\,\d W_\zs{t}\,,\quad t\ge 0\,,\quad X^*_\zs{0}\,=\,x\,,
\end{equation}
where $a^*(t,x)=a(t,x,\vartheta^*(t,x))$ and
$b^*(t,x)=b(t,x,\vartheta^*(t,x))$.
Moreover, the optimal control process
$\varsigma^*_\zs{t}=\vartheta^*(t,X^*_\zs{t})$ for $0\le t\le T$ belongs to $\cV$.}\\[3mm]
$\H_\zs{4})$ {\em There exists some $\delta>1$ such that
for all $0\le t\le T$ and $x>0$
$$
\E_\zs{t,x}\,
\sup_{t\le s\le T}\,(z(s,X^*_\zs{s}))^\delta\,
<\,\infty\,.
$$
}

\begin{theorem}\label{Th.A.1}
Assume that $\cV\ne\emptyset$ and $\H_\zs{1}-\H_\zs{4}$ hold.
Then for all $t\in [0,T]$ and for all $x> 0$ the solution to the
Hamilton-Jacobi-Bellman equation
(\ref{A.6}) coincides with the optimal value of the cost function,
i.e. $z(t,x)=J^*(t,x)=J^*(t,x,\varsigma^*)$, where the optimal strategy
$\varsigma^*$ is defined in  $\H_\zs{2}$ and $\H_\zs{3}$.
\end{theorem}

\noindent {\bf Proof.}
For $\varsigma\in\cV$ let $X^{\varsigma}$ be the associated wealth
process with initial value $X^{\varsigma}_0=x$.
Define stopping times
$$
\tau_\zs{n}\,=\,\inf\left\{s\ge\,t\,:\,
\int^s_\zs{t}\,b^2(u,X^{\varsigma}_\zs{u},\varsigma_\zs{u})\,z^2_\zs{x}(u,X^{\varsigma}_\zs{u})\,
\d u\,\ge\,n\right\}\wedge T\,.
$$
Note that condition \eqref{A.2} implies that $\tau_\zs{n}\to T$ as $n\to\infty$ a.s..
By  continuity of  $z(\cdot,\cdot)$ and of $(X^{\varsigma}_\zs{t})_{0\le t\le T}$
we obtain
\begin{equation}\label{A.10}
\lim_\zs{n\to\infty}\,z(\tau_\zs{n},X^{\varsigma}_\zs{\tau_\zs{n}})\,=\,
z(T,X^{\varsigma}_\zs{T})\,=\,h(X^{\varsigma}_\zs{T})\ \ \ \ \ \ \mbox{a.s..}
\end{equation}
Theorem~\ref{Th.A.2} guarantees that we can invoke It\^o's formula, and
we conclude from \eqref{A.1}
\begin{align}  \nonumber
 z(t,x) &= \,\int^{\tau_\zs{n}}_\zs{t}\,f(s,X^{\varsigma}_\zs{s},\varsigma_\zs{s})\,\d s\,+
\,z(\tau_\zs{n},X^{\varsigma}_\zs{\tau_\zs{n}})
-\int^{\tau_\zs{n}}_\zs{t}\,(z_\zs{t}(s,X^{\varsigma}_\zs{s})\\
&+
\,H_\zs{1}(s,X^{\varsigma}_\zs{s},\varsigma_\zs{s}))\,\d s
-\,\int^{\tau_\zs{n}}_\zs{t}\,
b(u,X^{\varsigma}_\zs{u},\varsigma_\zs{u})\,z_\zs{x}(u,X^{\varsigma}_\zs{u})\,\d W_\zs{u}\,,  \label{A.11}
\end{align}
where
$$
H_\zs{1}(s,x,\vartheta)\,=\,
H_\zs{0}(t,x,z_\zs{x}(t,x),z_\zs{xx}(t,x),\vartheta)\,.
$$
Condition $\H_\zs{1}$ implies
$$
z(t,x)
\ge
\E_\zs{t,x}\,\int^{\tau_\zs{n}}_\zs{t}\,f(s,X^{\varsigma}_\zs{s},\varsigma_\zs{s})\,
\d s\,+
\,\E_\zs{t,x}z(\tau_\zs{n},X^{\varsigma}_\zs{\tau_\zs{n}})\,.
$$
Moreover,  by monotone convergence for the first term and
 Fatou's lemma for the second, and by observing \eqref{A.10} we obtain
\begin{align}\nonumber
\lim_\zs{n\to\infty}&
\E_\zs{t,x}\,\int^{\tau_\zs{n}}_\zs{t}\,f(s,X^{\varsigma}_\zs{s},\varsigma_\zs{s})\,
\d s\,+ \,\lim_\zs{n\to\infty}
\E_\zs{t,x}z(\tau_\zs{n},X^{\varsigma}_\zs{\tau_\zs{n}})\\ \label{A.12}
&\ge
\E_\zs{t,x}\,\int^{T}_\zs{t}\,f(s,X^{\varsigma}_\zs{s},\varsigma_\zs{s})\,
\d s\,+
\,\E_\zs{t,x}\,h(X^{\varsigma}_\zs{T}):= J(t,x,\varsigma)\,,\quad 0\le t\le T\,.
\end{align}
Therefore, $z(t,x)\ge J^*(t,x)$ for all $0\le t\le T$.\\[2mm]
Similarly, replacing $\varsigma$ in \eqref{A.11} by $\varsigma^*$ as defined by $\H_\zs{2}-\H_\zs{3}$ we obtain
$$
 z(t,x)\,=\,\E_\zs{t,x}\,\int^{\tau_\zs{n}}_\zs{t}\,f(s,X^*_\zs{s},\varsigma^*_\zs{s})\,
\d s\,+\,\E_\zs{t,x}\,z(\tau_\zs{n},X^*_\zs{\tau_\zs{n}}) \,.
$$
Condition $\H_\zs{4}$ implies that the sequence
 $(z(\tau_\zs{n},X^*_\zs{\tau_\zs{n}}))_{n\in\bbn}$ is uniformly integrable.
 Therefore, by \eqref{A.10},
$$
\lim_{n\to\infty}\,\E_\zs{t,x}\,z(\tau_\zs{n},X^*_\zs{\tau_\zs{n}}) \,
=\,\E_\zs{t,x}\,
\lim_{n\to\infty}\,z(\tau_\zs{n},X^*_\zs{\tau_\zs{n}})\,=\,\E_\zs{t,x}\,h(X^*_\zs{T})\,,
$$
and we obtain
\begin{align*}
 z(t,x)\,&=\,\lim_\zs{n\to\infty}
\E_\zs{t,x}\,\int^{\tau_\zs{n}}_\zs{t}\,f(s,X^*_\zs{s},\varsigma^*_\zs{s})\,
\d s\,+\,\lim_{n\to\infty}\E_\zs{t,x}\,z(\tau_\zs{n},X^*_\zs{\tau_\zs{n}}) \\
&=\,
\E_\zs{t,x}\,\left(\int^{T}_\zs{t}\,f(s,X^*_\zs{s},\varsigma^*_\zs{s})\,
\d s\,+\,h(X^*_\zs{T})\right)\\
&=\,J(t,x,\varsigma^*)\,.
\end{align*}
Together with (\ref{A.12}) we arrive at $z(t,x)=J^*(t,x)$.
This proves Theorem~\ref{Th.A.1}.
\eproof

\begin{remark}\label{Re.A.2}\rm
Note that in contrast to the usual verification theorem
(see e.g. Touzi~\cite{To}, Theorem 1.4) we do not assume that
equation \eqref{A.6} has a solution for all $t\in [0,T]$, but only
for almost all $t\in [0,T]$. This provides the possibility to consider
 market models as in \eqref{2.1} with discontinuous functional coefficients.
Moreover, in the usual verification theorem the
function $f(t,x,\vartheta)$ is bounded with respect to $\vartheta\in\cK$
or integrable with all moments finite.
This is an essential difference of our situation as for the optimal
 consumption problem
$f$ is not bounded over $\vartheta\in\cK$ and we do not assume that $f$  is integrable.
\halmos\end{remark}

\subsection{A Special Version of It\^o's Formula}
\label{subsec:A.4}

  We prove It\^o's formula for functions satisfying $\H_\zs{1}$, an extension,
which to the best of our knowledge can not be found in the literature.
Consider the It\^o equation
\beao\label{A.13}
\d \xi_\zs{t}=a_t\,\d t\,+\,b_t\,\d W_t\,,
\eeao
where the stochastic processes
$a=(a_t)_\zs{0\le t\le T}$ and $b=(b_t)_\zs{0\le t\le T}$ are measurable, adapted and satisfy
for the investment horizon $T>0$
\begin{equation}\label{A.14}
\int^T_\zs{0}\,(|a_\zs{t}| + b^2_\zs{t})\,\d t\,<\,\infty\quad\mbox{a.s..}
\end{equation}

\begin{theorem}\label{Th.A.2}
Let $f\,:\,[0,T]\times (0,\infty)\to [0,\infty)$  satisfy $\H_\zs{1}$.
Assume that the process $\xi$ is a.s. positive on $0\le t\le T$.
Then $(f(t,\xi_t))_{0\le t\le T}$ is the solution to
\begin{equation}\label{A.15}
\d f(t,\xi_t)=\big(f_\zs{t}(t,\xi_t)+ f_\zs{x}(t,\xi_t)a_t+
\frac{1}{2}f_\zs{xx}(t,\xi_t)\big)b^2_t\,\d t+
f_\zs{x}(t,\xi_t)b_t\,\d W_t\,.
\end{equation}
\end{theorem}

\begin{remark}\label{Re.A.3}\rm
Note that in contrast to the usual It\^o formula
we do not assume that $f$ has a continuous derivative with respect to
$t$ and continuous derivatives with respect to $x$ on the whole of $\bbr$.
For example, the function \eqref{4.3} for
$\gamma_\zs{1}=\gamma_\zs{2}=\gamma\in (0,1)$
factorises into $z(t,x)=Z(t)x^\gamma$, i.e.
is not continuosly differentiable with respect to $x$ on $\bbr$.
\halmos\end{remark}

\proof
First we prove \eqref{A.15} for bounded processes $a$ and $b$, i.e.
we assume that for some constant $L>0$
\begin{equation}\label{A.16}
\sup_\zs{0\le t\le T}(|a_t|+|b_t|)\le L\quad\quad\mbox{a.s.}\,.
\end{equation}
Let $(t_k)_\zs{1\le k\le n}$ be a partition of $[0,T]$, more precisely,
take $t_k=kT/n$, and consider the telescopic sums
\beao
f(T,\xi_\zs{T})-f(0,\xi_0)&=& \sum^n_\zs{k=1}\,(f(t_k,\xi_\zs{t_k})-f(t_{k-1},\xi_\zs{t_{k}}))\\
&& +\, \sum^n_\zs{k=1}\,(f(t_\zs{k-1},\xi_\zs{t_k})-f(t_{k-1},\xi_\zs{t_{k-1}}))\\
&:=& \sum_\zs{1,n}+\sum_\zs{2,n}\,.
\eeao
Taking condition \eqref{A.7} into account we can represent the first sum as
\beao
\Sigma_\zs{1,n} &=&
\sum^n_\zs{k=1}\,\int^{t_k}_\zs{t_{k-1}}\,f_\zs{t}(u,\xi_\zs{t_k})\,\d u
\, = \, \int^{T}_\zs{0}\,f_\zs{t}(u,\xi_\zs{u})\,\d u+r_\zs{1,n}\,,
\eeao
where
$$
r_\zs{1,n}=\sum^n_\zs{k=1}\,\int^{t_k}_\zs{t_{k-1}}\,
(f_\zs{t}(u,\xi_\zs{t_k})- f_\zs{t}(u,\xi_\zs{u}))\d u\,.
$$
Now we prove that $r_\zs{1,n}\stp 0$ as $n\to\infty$.
To this end we introduce the stopping time,
\begin{equation}\label{A.17}
\tau_\zs{N}=\inf\{t\ge 0\,:\,\xi_t+\xi^{-1}_t\ge N\}\wedge T\,,\quad N>0\,.
\end{equation}
As the process $\xi$ is continuous and a.s. positive,
\begin{equation}\label{A.18}
\lim_\zs{N\to\infty}\P(\tau_\zs{N}<T)=0\,,
\end{equation}
and, hence, $\tau_N\stp T$ as $N\to\infty$.
Moreover, the modulus of continuity of the process $\xi$ satisfies
\begin{equation}\label{A.19}
\Delta_\zs{\epsilon}(\xi,[0,T])\, := \,
\sup_\zs{|t-s|\le \epsilon\,,\,s,t\in [0,T]}
|\xi_t-\xi_s| \,\stas \,0\,,\quad \epsilon\to 0\,.
\end{equation}
Note now that condition \eqref{A.8} implies that for every $N>1$
$$
F^*(\eta,N) :=\int^{T}_\zs{0}\,\sup_\zs{x,y,\in K_\zs{N}\,,\,|x-y|<\eta}\,
|f_\zs{t}(u,x)- f_\zs{t}(u,y)|\d u \to 0
\quad\mbox{as}\quad \eta\to 0\,,
$$
where $K_\zs{N}=[N^{-1},N]$.
This implies that for every $\delta>0$ there exists
$\eta_\zs{\delta}>0$ such that $F^*(\eta_\zs{\delta},N)<\delta$.
Moreover, taking into account that for $\epsilon=T/n$
the random variable $r_\zs{1,n}$ is bounded on the $\omega$-set
$$
\{\Delta_\zs{\epsilon}(\xi,[0,T])\le\eta_\zs{\delta}\}\cap\{\tau_\zs{N}=T\}
$$
by $|r_\zs{1,n}|\le F^*(\eta_\zs{\delta},N)<\delta$,
we obtain that
$$
\P(|r_\zs{1,n}|>\delta)\le \P(\Delta_\zs{\epsilon}(\xi,[0,T])>\eta_\zs{\delta})+
\P(\tau_\zs{N}<T)\,.
$$
Relations \eqref{A.18} and \eqref{A.19} imply $r_\zs{1,n}\stp 0$ as $\nto$.
Now define
$$
r_\zs{2,n} :=\Sigma_\zs{2,n}-\int^T_\zs{0}f_\zs{x}(t,\xi_t)\d \xi_t-
\frac{1}{2}\int^T_\zs{0}f_\zs{xx}(t,\xi_t)\,b^2_\zs{t}\d t\,.
$$
We show that $r_\zs{2,n}\,\stp\,0$ as $\nto$.
A Taylor expansion gives
\beam\label{A.20}
\Sigma_\zs{2,n}&=&\sum^n_\zs{k=1}\,f_\zs{x}(t_\zs{k-1},\xi_\zs{t_\zs{k-1}})
\Delta\xi_\zs{t_k}\,+\,
\frac{1}{2}\sum^n_\zs{k=1}\,f_\zs{xx}(t_\zs{k-1},\xi_\zs{t_\zs{k-1}})
\int^{t_k}_\zs{t_\zs{k-1}}b^2_u\d u\nonumber\\
&&+\,\frac{1}{2}\sum^n_\zs{k=1}
\,f_\zs{xx}(t_\zs{k-1},\xi_\zs{t_\zs{k-1}})\alpha_\zs{k}
+\frac{1}{2}\sum^n_\zs{k=1}\,\wh{f}_\zs{k}(\Delta\xi_\zs{t_k})^2\,,
\eeam
where $\alpha_\zs{k}=(\Delta\xi_\zs{t_k})^2-\int^{t_k}_\zs{t_\zs{k-1}}b^2_u\d u$,
$\wh{f}_\zs{k}=f_\zs{xx}(t_\zs{k-1},\wh{\xi}_\zs{t_k})
-f_\zs{xx}(t_\zs{k-1},\xi_\zs{t_\zs{k-1}})$ and   \\
$ \wh{\xi}_\zs{t_k}=\xi_\zs{t_\zs{k-1}}+\theta_\zs{k}\Delta\xi_\zs{t_k}$
with $\theta_\zs{k}\in [0,1]$.
Now taking into account that as $\nto$
\beao
\sum^n_\zs{k=1}\,f_\zs{x}(t_\zs{k-1},\xi_\zs{t_\zs{k-1}})
\Delta\xi_\zs{t_k} & \stp & \int^T_\zs{0}f_\zs{x}(t,\xi_t)\d \xi_t \\
\sum^n_\zs{k=1}\,f_\zs{xx}(t_\zs{k-1},\xi_\zs{t_\zs{k-1}})
\int^{t_k}_\zs{t_\zs{k-1}}b^2_u\d u
& \stas &
\int^T_\zs{0}f_\zs{xx}(t,\xi_t)\,b^2_\zs{t}\d t
\eeao
it suffices to show that the last two terms in \eqref{A.20} tend to zero in probability.
To this end we represent the first sum as
$$
\sum^n_\zs{k=1}\,f_\zs{xx}(t_\zs{k-1},\xi_\zs{t_\zs{k-1}})\alpha_\zs{k}=
M_\zs{n}+R_\zs{n}\,,
$$
where
\beao
M_\zs{n} &=&  \sum^n_\zs{k=1}\,f_\zs{xx}(t_\zs{k-1},\xi_\zs{t_\zs{k-1}})\eta_\zs{k}
\quad\mbox{with}\quad
\eta_\zs{k}=(\int^{t_k}_\zs{t_{k-1}}b_u\d W_\zs{u})^2-
\int^{t_k}_\zs{t_\zs{k-1}}b^2_u\d u\,,\\
R_n &=& \sum^n_\zs{k=1}\,f_\zs{xx}(t_\zs{k-1},\xi_\zs{t_\zs{k-1}})\alpha^*_k
\quad\mbox{with}\quad
\alpha^*_k=(\Delta\xi_\zs{t_k})^2-(\int^{t_k}_\zs{t_{k-1}}b_u\d W_\zs{u})^2\,.
\eeao
First we estimate the martingale part in this representation.
Note that on the set $\{\tau_\zs{N}=T\}$ the martingale part
coincides with the bounded martingale
$$
M_n=
\sum^n_\zs{k=1}f_\zs{xx}(t_\zs{k-1},\xi_\zs{t_\zs{k-1}\wedge \tau_\zs{N}})\eta_\zs{k}\,.
$$
Taking into account  that
$$
|f_\zs{xx}(t_\zs{k-1},\xi_\zs{t_\zs{k-1}\wedge \tau_\zs{N}})|\le
\sup_\zs{t\in [0,T]\,,y\,\in [N^{-1},N]}|f_\zs{xx}(t,y)|:=M_*
$$
we obtain
\begin{align*}
\E {M}^2_n &=
\E \sum^n_\zs{k=1}f^2_\zs{xx}(t_\zs{k-1},{\xi}_\zs{t_\zs{k-1}\wedge \tau_\zs{N}})
\eta^2_\zs{k}
\le \, M^2_*\sum^n_\zs{k=1}
\E\,\left(\int^{t_k}_\zs{t_{k-1}}b_u\d W_\zs{u}\right)^4\\
&\le 3L^4M^2_*\sum^n_\zs{k=1}(\Delta t_\zs{k})^2
 =  3L^4M^2_*T^2\frac{1}{n} \quad\to \, 0\,,\quad n\to\infty\,.
\end{align*}
In the last inequality we used the bound \eqref{A.16} for $b$.
We conclude
\begin{equation}\label{A.21}
M_n \stp 0\,,\quad\nto\,.
\end{equation}
Using the convergence \eqref{A.19} also for
$I(t)=\int^{t}_\zs{0}b_u\d W_\zs{u}$ and the upper bound \eqref{A.16} for $a$
we obtain
\begin{align*}
|\alpha^*_k|&\le \Big(\int^{t_k}_\zs{t_\zs{k-1}}a_u\d u\Big)^2+
2\int^{t_k}_\zs{t_\zs{k-1}}|a_u|\d u
\Big|\int_{t_\zs{k-1}}^{t_k}b_u\d W_\zs{u}\Big|\\
&\le L^2 (\Delta t_\zs{k})^2+
2 L\Delta_\zs{\epsilon}(I,[0,T])\,\Delta t_\zs{k}\,,
\end{align*}
where $\epsilon=\Delta t_\zs{k}=T/n$. This yields
$\lim_\zs{n\to\infty}\sum^n_\zs{k=1}|\alpha^*_k|=0$ a.s.
We use analogous arguments as for \eqref{A.21}
to show that $R_n\stp 0$.
Taking also into account that
$\sum^n_\zs{k=1}(\Delta \xi_\zs{t_k})^2$ is bounded in probability, i.e.
$$
\lim_\zs{m\to\infty}
\P\left(\sum^n_\zs{k=1}(\Delta \xi_\zs{t_k})^2 \ge m\right)=0\,,
$$
it is easy to see that the last sum in \eqref{A.20} tends to zero
in probability. This proves Ito's formula \eqref{A.15}
for bounded coefficients $(a_\zs{t})$ and $(b_\zs{t})$.\\[2mm]
To prove Ito's formula under condition \eqref{A.14}
we introduce for $L\in\bbn$ the sequence of
processes $(\xi^{L}_\zs{t})_\zs{0\le t\le T}$ by
$$
\d \xi^{L}_\zs{t}=a^{L}_t\,\d t\,+\,b^{L}_t\,\d W_t\,,\quad \xi^{L}_\zs{0}=\xi_0\,,
$$
where $a^{L}_t:=a_t\chi_\zs{\{|a_t|\le L\}}$ and
$b^{L}_t:=b_t\chi_\zs{\{|b_t|\le L\}}$.
For each of these processes we already proved \eqref{A.15}.
Therefore we can write
\begin{equation}\label{A.22}
f(T,\xi^{L}_\zs{T})=f(0,\xi_\zs{0})+\int^T_\zs{0}A^{L}_\zs{t}\,\d t+
\int^T_\zs{0}B^{L}_\zs{t}\,\d W_\zs{t}\,,
\end{equation}
where
$A^{L}_\zs{t}=f_\zs{t}(t,\xi^{L}_t)+f_\zs{x}(t,\xi^{L}_t)a^{L}_\zs{t}
+f_\zs{xx}(t,\xi^{L}_t)(b^L_\zs{t})^2/2$
and $B^{L}_\zs{t}=f_\zs{x}(t,\xi^{L}_t)b^{L}_\zs{t}$.
Note that \eqref{A.14} implies immediately
$$
\lim_\zs{L\to\infty}\int^T_\zs{0}(|a^{L}_t-a_t|+(b^{L}_t-b_t)^2)\,\d t=0
\quad\mbox{a.s.}\,.
$$
Taking this into account we show that
\begin{equation}\label{A.23}
\sup_\zs{0\le t\le T}|\xi^{L}_\zs{t}-\xi_\zs{t}|\stp 0\,,\quad L\to\infty\,.
\end{equation}
Indeed, from the definitions of $\xi$ and $\xi^{L}$ we obtain that
$$
\sup_\zs{0\le t\le T}|\xi^{L}_\zs{t}-\xi_\zs{t}|\le
\int^T_\zs{0}|a^{L}_t-a_t|\d t+\sup_\zs{0\le t\le T}
\Big|\int^t_\zs{0}(b^{L}_t-b_t)\,\d W_\zs{t}\Big|\,.
$$
Thus for \eqref{A.23} it suffices to show that the last term
in this inequality tends to zero as $L\to\infty$.
By Lemma 4.6, p. 102 in Liptser and Shiryaev~\cite{LiSh})
we obtain for every $\epsilon>0$
$$
\P\left(
\sup_\zs{0\le t\le T}
\left|\int^t_\zs{0}(b^{L}_t-b_t)\,\d W_\zs{t}\right|\,\ge\,\delta
\right)\le \frac{\epsilon}{\delta^2}+
\P\left(
\int^T_\zs{0}(b^{L}_t-b_t)^2\d t\,\ge\,\epsilon
\right)\,.
$$
This implies \eqref{A.23}.
Taking now the limit in \eqref{A.22} for $L$ to infinity we obtain \eqref{A.15}. \eproof


\begin{thebibliography}{99.}%

\bibitem{ArDeEbHe}
Artzner, P., Delbaen, F., Eber, J.-M., Heath, D.:
Coherent measures of risk.  {\em Math. Finance.},
\textbf{9}, 203--228  (1999)

\bibitem{BaSha}
Basak, S., Shapiro, A. :
Value at Risk based risk management: optimal policies and asset prices.
 {\em Review of Financial Studies.}, \textbf{14}, 371--405  (1999)

\bibitem{GaGrWu}
Gabih, A., Grecksch, W., Wunderlich, R. :
Dynamic portfolio optimization with bounded shortfall risks.
{\em Stoch. Anal. Appl.}, \textbf{23}, 579--594  (2005)

\bibitem{EmKlKo}
Emmer, S., Kl\"uppelberg, C., Korn, R. :
Optimal portfolios with bounded Capital-at-Risk.
{\em Math. Finance.}, \textbf{11}, 365--384  (2001)

\bibitem{Jo}
Jorion, P. : {\em Value at Risk}.
McGraw-Hill, New York (2001)

\bibitem{KaSh1}
Karatzas, I. and Shreve, S.E. :
{\em Brownian Motion and Stochastic Calculus.}
Springer, Berlin  (1988)

\bibitem{KaSh2}
Karatzas, I. and Shreve, S.E. :
{\em Methods of Mathematical Finance.}
Springer, Berlin  (2001)


\bibitem{Ko}
Korn, R. :
{\em Optimal Portfolios.}
World Scientific, Singapore (1997)

\bibitem{LiSh}
Liptser, R.S., Shirayev, A.N. :
{\em Statistics of Random Processes I. General Theory.}
Springer, New York (1977)

\bibitem{Me}
Merton, R.C. :
{\em Continuous-Time Finance.}
Blackwell, Cambridge MA (1990)

\bibitem{To}
Touzi, N. : Stochastic Control Problems, Viscosity Solutions and Applications to Finance.
{\em Publications of the Scuola Normale Superiore of Pisa},
Scuola Normale Superiore, Pisa (2004)


\end{thebibliography}
\end{document}